\documentclass[times]{MyArticle}

\usepackage[colorlinks,bookmarksopen,bookmarksnumbered,citecolor=blue,urlcolor=blue]{hyperref}

\usepackage[pgfplots,usenames,dvipsnames,svgnames,table]{xcolor}
\usepackage{tabularx}
\usepackage{pgfplots}
\pgfplotsset{width=4cm,compat=1.15}
\usepackage{caption}
\usepackage{subcaption}
\usepackage{siunitx}
\let\qty\SI
\let\unit\si

\usepackage{mathtools}
\usepackage{textcomp}

\begin{document}
\title{Meshfree One-Fluid Modelling of Liquid-Vapor Phase Transitions}

\author{%
Pratik Suchde\affil{1,2}\corrauth,
Heinrich Kraus\affil{1,3}, %
Benjamin Bock-Marbach \affil{1}, %
J{\"o}rg Kuhnert\affil{1} %
}

\address{\affilnum{1}Fraunhofer ITWM, Fraunhofer-Platz 1, 67663 Kaiserslautern, Germany. \break %
\affilnum{2}University of Luxembourg, 2 avenue de l'université, L-4365 Esch-sur-alzette, Luxembourg \break%
\affilnum{3} Institut für Mathematik, Universität Kassel, Heinrich-Plett-Straße 40, 34132 Kassel, Germany
}

\corraddr{E-mail: pratik.suchde@gmail.com, pratik.suchde@uni.lu}

\begin{abstract}
We introduce a meshfree collocation framework to model the phase change from liquid to vapor at or above the boiling point. While typical vaporization or boiling simulations focus on the vaporization from the bulk of the fluid, here we include the possibility of vaporization from the free surface, when a moving fluid comes into contact with a superheated surface. We present a continuum, one-fluid approach in which the liquid and vapor phases are modeled with the same constitutive equations, with different material properties. The novelty here is a monolithic approach without explicit modeling of the interface between the phases, neither in a sharp nor diffuse sense. Furthermore, no interface boundary conditions or source terms are needed between the liquid and vapor phases. Instead, the phase transition is modeled only using material properties varying with temperature. Towards this end, we also present an enrichment of strong form meshfree generalized finite difference methods (GFDM) to accurately capture derivatives in the presence of jumps in density, viscosity, and other physical properties. The numerical results show a good agreement with experimental results, and highlight the ability of our proposed framework to model phase changes with large jumps. 
\end{abstract}


\maketitle


\section{Introduction}
\label{sec:Introduction}

The process of phase transition or phase change occurs everywhere around us, both in nature and in industrial processes. They have been widely studied both theoretically and experimentally. While computer simulations of phase transitions have become increasingly popular in the last few decades, there remain many challenges in modeling and simulating these processes accurately. 

In the present work, we restrict our discussion to so-called \emph{first-order} phase transitions that involve latent heat. Several different approaches have been considered for simulating such phase change processes. On one hand, microscopic simulations have been widely considered \cite{dePablo1999}, which simulate the process at a molecular or atomic level. These methods can be classified into one of two approaches: (i) molecular dynamics (MD) simulations which integrate equations of motions on interacting particles, see, for example, \cite{Duncan2011, Matsumoto1998, Nguyen2017, Xu2004}, and (ii) Monte Carlo (MC) methods which rely on stochastic processes, 
see, for example, \cite{Binder1992,  Satoh2016, Wang2010, Wilding2001}. While these microscopic phase change models can be very physically accurate, they are only suitable for simulations over very small length scales, typically smaller than millimeters. 

On the other hand, continuum approaches are used to model the phase change process on a macroscopic level. These models are more appropriate for large length scale simulations of phase transitions. The primary difficulties in these models arise due to the volume expansion or contraction, and the evolving phase boundaries that undergo large deformations and possibly even topological changes. Based on similar notions in classical multiphase flow problems \cite{Lee2012}, continuum phase change simulations are usually classified into two categories based on how the interface between the phases is treated: (i) The classical approach is a sharp interface method, which assumes a clear and distinct interface of zero thickness between the two phases. The phase transition is modeled as interface boundary conditions for heat and mass transfer, see for example, \cite{Can2012, Gibou2007, Welch2000}. (ii) An alternative approach that has become very popular is the diffusive interface approach, that postulates an interface of a small non-zero thickness. Here, the interface conditions are modeled as source terms in the governing equations across the entire thickness of the diffuse interface, see for example, \cite{Gomez2019, Pan2019, Samkhaniani2017, Sun2004}. Mesoscopic approaches for phase change modeling have also been proposed, and typically use the Lattice Boltzmann method (LBM) \cite{Li2016}.

However, for many practical applications of phase transitions, none of the above methods are suitable. The motivation behind the present work is the vaporization of cutting fluid during metalworking processes. 
In these processes, chips are removed from a metal workpiece, using a cutting tool, to obtain the desired shape of the workpiece. These processes generate a large amount of heat due to friction and metal deformation \cite{Uhlmann2013}. Thus, a cutting fluid, also referred to as coolant or lubricant, is used to rapidly remove heat without needing to pause the production process, while also lubricating and preventing unwanted thermal expansion \cite{Uhlmann2021b, Uhlmann2021a}. The temperatures at the cutting region can often be much higher than the boiling temperature of the cutting fluid. As a result, as the fluid comes into contact with the workpiece or the tool in the cutting zone, it instantly vaporizes. Thus, to accurately simulate such cutting processes, the liquid-vapor transition must be modeled. However, existing phase change models are unable to represent this complex process. Firstly, microscopic models are infeasible for such large scale simulations. A macroscopic sharp interface method is not applicable due to the complicated 3D interface with changing topology between the liquid and vapor phases. Moreover, unlike typical macroscopic phase change simulations where the only free boundary is the interface between the phases, here both the liquid and the gas themselves have their own external free surfaces too. In fact, rapid boiling of the liquid occurs at and near a free surface. Diffuse interface methods would fail near the region where multiple interfaces intersect: the liquid-vapor interface, external free surfaces for both liquid and vapor phases, and solid-liquid and solid-vapor interfaces where the cutting fluid hits the workpiece and tool. 


Meshfree methods have become a popular alternative to conventional mesh-based simulation methods, especially for fluid flow applications. Their advantages become the most relevant for applications with complex geometries, moving interfaces, and free boundaries, each of which holds true for our vaporization of cutting fluid application. Another advantage of a meshfree method is the ease of incorporating moving Lagrangian frameworks. While Lagrangian frameworks are considered to be very accurate in representing moving boundaries and interfaces, their use with mesh-based methods is limited to applications with small deformations \cite{Lee2012}. This is primarily because mesh deformation as a result of moving the mesh requires an expensive remeshing step. In contrast, in meshfree methods point cloud deformation can be fixed locally and is thus much cheaper. As a result, meshfree Lagrangian methods are especially suited for applications with large deformations and topology changes of interfaces. We thus choose a meshfree approach to model the application at hand.

In this paper, we introduce a novel meshfree Lagrangian framework to simulate phase change processes. Motivated by similar ideas in cavitation modeling \cite{Goncalves2009}, we use a one-fluid or homogeneous approach, in which both phases of the phase transition process are represented by a single fluid, with one set of conservation equations and varying material properties representing the different phases. We note that the so-called front tracking method for phase change \cite{Juric1998} also uses a one-fluid or a ``one-field" approach. However, they explicitly track the interface and use two different grids for the flow equations and the interface. In contrast, here, we do not track the interface at all, and thus only one grid (more appropriately, one point cloud) is needed. Mass and energy transfer between the phases are represented directly by the particle-based Lagrangian nature of the discretization. This has the advantage of not needing to explicitly track or capture the interface, nor needing to determine and impose any interfacial conditions, which could have been cumbersome for interfaces with very complex topology. Using such an approach in a fully meshfree setting raises several challenges, which are highlighted and addressed in the present work. We introduce modifications to collocation-based meshfree derivative computation to enable accurate capturing of the sudden changes in the material properties. Furthermore, we introduce a framework for modeling the varying material properties that helps in attaining numerical stability while maintaining physical consistency. 


Given the aforementioned motivation, we focus on liquid-to-vapor phase change throughout the present work. However, the methods introduced here could be directly used to model any first-order phase transition. 

This paper is organized as follows. We first introduce the constitutive equations underlying the one-fluid model in Section~\ref{sec:Equations}, followed by explaining how the phase change is modeled by varying material properties in Section~\ref{sec:Material}. The meshfree discretization procedure and enrichment process are then explained in Section~\ref{sec:Meshfree}. Section~\ref{sec:TimeInteg} describes some details of the numerical scheme, and a short comparison with other phase change models is presented in Section~\ref{sec:Comparisons}. Numerical applications of the method are explained in Section~\ref{sec:Simulations}, followed by brief concluding remarks in Section~\ref{sec:Conclusion}.


\section{Constitutive Equations}
\label{sec:Equations}

The underlying constitutive equations used to model both the liquid and the vapor phase as a single one-fluid model are the standard conservation equations of mass, momentum, and energy, written in a Lagrangian formulation
\begin{subequations} \label{Eq:CoEquations}
\begin{align}
	\frac{D \rho}{Dt} &= -\rho \nabla \cdot \vec{v}, \label{Eq:CoMass}\\ 
	\frac{D  \vec{v}  }{Dt} &= \frac{1}{\rho} \nabla \cdot \mathbf{S} - \frac{1}{\rho} \nabla p 
	+  \vec{g}, \label{Eq:CoMomentum}\\
	\rho \frac{D  E }{Dt} &= \nabla \cdot (\mathbf{S}\vec{v}) 
	- \nabla \cdot ( p \vec{v} ) + 
	\rho \vec{g} \cdot \vec{v} +
	\nabla \cdot \left( \lambda_{\text{eff}} \nabla T \right) + q, \label{Eq:CoEnergy}
\end{align}
\end{subequations}
where the energy is given by $E = c_v T + \frac{1}{2}\vec{v} \cdot \vec{v}$. Furthermore, $\rho$ is the density, $\vec{v}$ is the velocity, $p$ is the pressure, $T$ is the temperature, while $t$ is the time, $\vec{g}$ is composed of both gravity and body forces, $c_v$ is the specific heat capacity, $\lambda_{\text{eff}}$ is the effective heat conductivity (see Section~\ref{sec:CVS}), and $q$ models the heat sources. The material or Lagrangian derivative is denoted by $\frac{D}{Dt}$. The stress tensor $\mathbf{S}$ can be composed of both viscous and non-viscous terms
\begin{equation}
	\mathbf{S}  = \mathbf{S}(\vec{v})
	= \mathbf{S}_{\text{visc}}(\vec{v}) + \mathbf{S}_{\text{solid}}(\vec{v}).
\end{equation}
The viscous stress tensor is given by 
\begin{equation}
	\mathbf{S}_{\text{visc}}(\vec{v}) 
	= \eta_{\text{eff}} \left(  
	\left( \nabla \vec{v} \right) + \left( \nabla \vec{v} \right)^T  - 
	\frac{2}{3} \nabla \cdot \vec{v} \, \mathbf{I} \right),
\end{equation}
where $\eta_{\text{eff}}$ denotes the effective viscosity  (see Section~\ref{sec:CVS}). The non-viscous parts of the stress tensor are grouped into the $\mathbf{S}_{\text{solid}}$ term governed by the appropriate material model. Since we are focusing on Newtonian liquid-to-vapor phase transition, we have $\mathbf{S}_{\text{solid}}=\textbf{0}$ in the applications considered. 

The governing equations \eqref{Eq:CoEquations} require a closure relation, which is given, depending on the application, either by an equation of state or by discrete curves determined by experiments. In each case, we have a relation $\rho = \rho(T, p)$, with the temperature and pressure both being functions of space $\vec{x}$ and time $t$. In many boiling simulations in literature, an assumption of the incompressibility of the vapor phase is often imposed (example, \cite{Esmaeeli2004, Samkhaniani2017, Welch2000}), to focus on the modelling of the phase transition process itself. Here, we allow the possibility of both incompressible and compressible vapor phases by using different closure relations. For actual applications, the choice of the appropriate closure model is a big challenge, as is also the case for phase change models which track the interface. 

The phase change from liquid to vapor is inherently a turbulent process. We thus incorporate a standard $k$-$\varepsilon$ turbulence model \cite{Launder1974, Versteeg2007} with fluctuating dilatation and source terms omitted. For more details on the turbulence model used, the parameters, and the numerical interpretation and integration of the turbulence model, we refer to our earlier work \cite{Michel2021}.

\section{Material properties}
\label{sec:Material}

Within the one-fluid model used here (see Section~\ref{sec:Equations}), the different phases are only distinguished by different material properties. This could lead to large jumps, or at minimum large gradients, in material properties where the phase transition occurs. We now introduce the numerical handling of these changing material properties such that both numerical stability and physical consistency are ensured. 

%
%
Here, the term \emph{interface} refers to the interface in the physical event of phase transition from liquid to vapor. We emphasize that this interface is not tracked explicitly in the numerical setting, neither in the sharp sense nor in a diffuse sense. Similarly, ``two phases" refers only to the physical liquid and vapor phases, while in the numerical setting, they are treated as a single phase computationally. 

\subsection{Boiling point}

An important question that arises in the one-fluid modeling of phase change concerns when the latent heat of the phase change must be taken into account. For complete physical accuracy, this should occur at a constant temperature. This would entail the latent heat of vaporization being modeled only at the normal boiling temperature or boiling point of the liquid. However, numerically, this can prove to be quite challenging, as it could mean reducing the simulation time step when a liquid nears the boiling point. To overcome this, we assume the phase change occurs over a small temperature range rather than modeling it at a single temperature. 

For a given pressure, the boiling point of the liquid phase is represented by $T_p^*$. Numerically, we spread this boiling point across a small interval of size $2\Delta T^*$. Thus, the vaporization process in our model occurs in the \emph{boiling temperature range} $[ T_p^* - \Delta T^*, T_p^* + \Delta T^*]$. The incorporation of the latent heat of vaporization is spread over this interval (more details in the next subsection). This provides a straightforward approach to modeling the phase change process without reducing the time step as the temperature approaches the phase change temperature. The use of a boiling temperature range here is similar in essence to the use of ``mushy zones" in solidification modelling \cite{Voller1987}. 

\subsection{Effective specific heat capacity}

In the boiling temperature range, specifically, in the semi-open intervals $[ T_p^* - \Delta T^*, T_p^* )$ and $(T_p^*,  T_p^* + \Delta T^*]$, our model must account for both the heat capacity of the material, as well as the latent heat of vaporization. For this, we introduce a numerical notion of \textit{heat capacity of vaporization}, referred to by $c_v^{\text{vaporization}}$. This will combine both the specific latent heat for the phase change process and the specific heat capacity to raise the temperature without phase change, into a single entity. As a result, the isochoric structure of the specific heat capacity is taken as
\begin{equation}
	c_v = \begin{cases}
				c_v^{\text{liquid}},       & \text{if }   T - T_p^*   < -\Delta T^*, \\
				c_v^{\text{vaporization}}, & \text{if } | T - T_p^* | <  \Delta T^*, \\
				c_v^{\text{vapor}},       & \text{if }   T - T_p^*   >  \Delta T^*,
	      \end{cases}
\end{equation}
%
where $c_v^{\text{liquid}}$ and $c_v^{\text{vapor}}$ are the specific heat capacities of the liquid and vapor phases, respectively. Both of these can be either constant or given by temperature-dependent curves. 

The value of $c_v^{\text{vaporization}}$ is given by ensuring physical consistency in the sense that the amount of heat required to take a unit mass of fluid from $T - T_p^*$ to $T + T_p^*$ should include the effects of both the specific latent heat and the specific heat capacity. Integrating the specific heat capacity over the phase transition region, we get
\begin{subequations}
\begin{align}
	\int_{T_p^* - \Delta T^*}^{T_p^* + \Delta T^*} c_v\, dT &= 
	\int_{T_p^* - \Delta T^*}^{T_p^* + \Delta T^*} c_v^{\text{vaporization}} \, dT \\
	&= c_v^{\text{liquid}} \Delta T^* + c_v^{\text{vapor}} \Delta T^* + \Delta H_p^{\text{vap}}, \label{Eq:cVapInt}
\end{align}
\end{subequations}
%
where $\Delta H_p^{\text{vap}}$ is the (specific) heat of vaporization. If $c_v^{\text{liquid}}$ or $c_v^{\text{vapor}}$ are not constant in this region, Eq.\,\eqref{Eq:cVapInt} can be changed accordingly with, for example, the term $c_v^{\text{liquid}} \Delta T^*$ replaced with $\int_{T_p^* - \Delta T^*}^{T_p^*} c_v^{\text{liquid}}\, dT$. For notational brevity, henceforth, we only consider the case of constant $c_v^{\text{liquid}}$ and $c_v^{\text{vapor}}$. 

Furthermore, we consider the numerical term $c_v^{\text{vaporization}}$ to also be constant, which leads to $\int_{T_p^* - \Delta T^*}^{T_p^* + \Delta T^*} c_v^{\text{vaporization}} \, dT  = c_v^{\text{vaporization}} 2 \Delta T^*$ and thus, the numerical notion of heat capacity of vaporization is given by 
\begin{equation}
	c_v^{\text{vaporization}} =  \frac{1}{2}( c_v^{\text{liquid}} + c_v^{\text{vapor}} ) + \frac{\Delta H_p^{\text{vap}}}{2\Delta T^*}.
\end{equation}

\subsection{Conductivity, viscosity, surface tension}
\label{sec:CVS}

The heat conductivity $\lambda$ is given by a heat conductivity for the liquid phase $\lambda^{\text{liquid}}$ at $T < T_p^* - \Delta T^*$, and a heat conductivity for the vapor phase $\lambda^{\text{vapor}}$ used at $T > T_p^* + \Delta T^*$. The heat conductivity in the boiling temperature region is taken to be a linear interpolation between these two values. The effective heat conductivity is taken as a sum of the laminar part $\lambda$ and the turbulent part, given by
\begin{equation}
		 \lambda^{\text{turbulent}} = 
	\left( c_{\mu} \frac{k^2}{\varepsilon} \right) \rho c_v.
\end{equation}

Similarly, the viscosity is given by the liquid phase viscosity $\eta^{\text{liquid}}$ and a vapor phase viscosity $\eta^{\text{vapor}}$, with a linear interpolation between the two in the boiling temperature zone. The effective viscosity is taken as a sum of the laminar part $\eta$ and the turbulent part, given by
\begin{equation}
	\eta^{\text{turbulent}} = 
	\left( c_{\mu} \frac{k^2}{\varepsilon} \right) \rho.
\end{equation}

Similarly, the surface tension is also given by a liquid phase and vapor phase surface tension with a linear interpolation in the boiling temperature zone, without any turbulent surface tension model. 

\subsection{Density}

As mentioned above, the density for both the liquid and vapor phase is given by the closure relation, depending on both pressure and temperature $\rho = \rho(T,p)$, where $T$ and $p$ are a function of both time $t$ and space $\vec{x}$. At a fixed pressure $p$, we use the notation $\rho(T,p) = \rho_p(T)$. 

The choice of the density in the phase change temperature region can have a significant impact on the numerical stability of the complete simulation, and thus plays a much larger role than that of the materials in Section~\ref{sec:CVS}. The simplest option is a linearly varying density in the phase transition range, as done in Section~\ref{sec:CVS}. Another option is a trigonometric ``smearing out", as is employed by many phase field methods for diffuse interface capturing \cite{Gomez2019}. Instead, we choose a weighted harmonic interpolation. 
For this, we first define the evaporation state variable
\begin{equation}
	x^{\text{vapor}} = \frac{T - (T_p^* - \Delta T^* )}{ 2 \Delta T^*}.
\end{equation}
Using this, and for a density $\rho^{\text{liquid}}$ at $T = T_p^* - \Delta T^*$, and $\rho_p^{\text{vapor}}$ (could be pressure dependent) at $T = T_p^* + \Delta T^*$, the density is determined by 
\begin{equation}
	\rho(T,p) = 
\begin{cases}
	\rho^{\text{liquid}},       & \text{if }   T - T_p^*   < -\Delta T^*, \\ 
	\left( \rho^{\text{liquid}}\rho_p^{\text{vapor}} \right) / \left( x^{\text{vapor}} \rho^{\text{liquid}} + (1-x^{\text{vapor}}) \rho_p^{\text{vapor}} \right),                     & \text{if } | T - T_p^* | <  \Delta T^*, \\
	\rho_p^{\text{vapor}},      & \text{if }   T - T_p^*   >  \Delta T^*.
\end{cases}
\end{equation}
Note that $\rho^{\text{liquid}}$ and $\rho_p^{\text{vapor}}$ can be temperature dependent.

\section{Meshfree discretization}
\label{sec:Meshfree}

In this section, we describe the meshfree point cloud based domain discretization, and the meshfree collocation approach to derivative computation. We then go on to introduce an enrichment procedure for derivative computation that is necessary to handle the large jumps and large gradients in the material properties.

\subsection{Meshfree point cloud based discretization}

\begin{figure}
  \centering
  \includegraphics[width=0.45\textwidth]{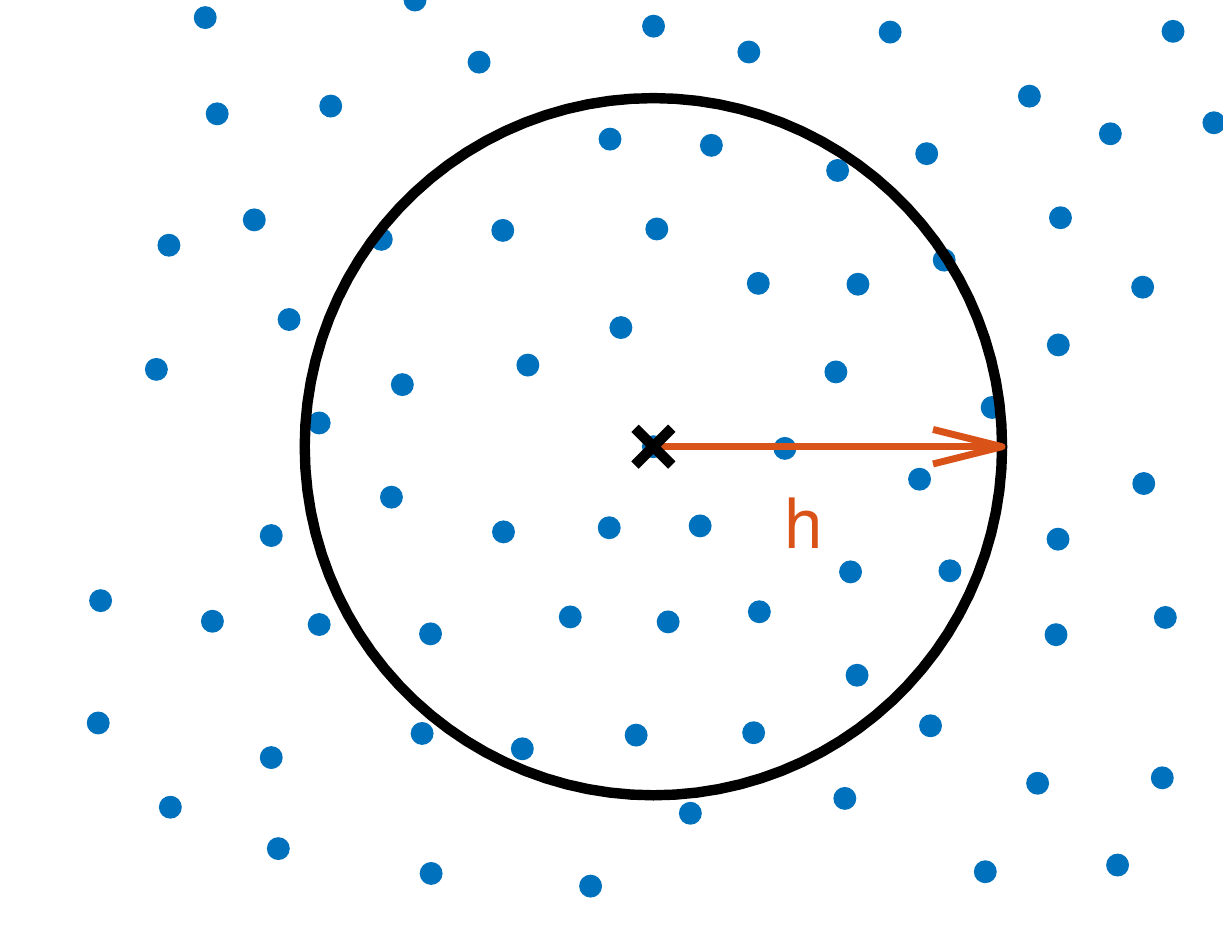}    
	\caption{A proximity-based neighborhood on a meshfree point cloud. For the point marked with the black cross, the black circle encloses the neighboring points over which all derivative approximations are carried out. }
  \label{Fig:PC}%
\end{figure}

The discretized computational domain $\Omega$, consisting of both the fluid and the vapor phases modeled as a single fluid, is composed of a cloud of $N = N(t)$ points. This includes points both in the interior of the domain and on all boundaries. The initial point cloud to discretize the domain is generated using a meshfree advancing front method \cite{Suchde2021_PCG}.
These points are approximation locations or collocation nodes, and not mass-carrying particles. Each point carries all physical and material properties of the fluid system. For every point $i = 1,2,\dots, N$, at location $\vec{x}_i$, numerical approximations are done on a set of neighboring points, referred to as its support or neighborhood
\begin{equation}
	S_i = \{ j \in \{1,2,\dots, N\} | \, \| \vec{x}_j - \vec{x}_i  \| \leq h_i  \},
\end{equation}
where $h_i = h(\vec{x}_i, t)$ is the smoothing length or interaction radius. Note that we have $i \in S_i$. An example of proximity-based neighborhoods is illustrated in Figure~\ref{Fig:PC}. 

\subsection{Adaptivity}

Since points are moving with the fluid velocity in a Lagrangian sense, the point cloud could become distorted. This would manifest in either points coming too close to each other, or hole formations in the point cloud where no points are present. In contrast to mesh distortion in mesh-based Lagrangian methods, this point cloud distortion can be easily fixed locally. This is done in two parts: (i) points that are closer than $r_{\text{min}}h$ apart are merged into a single point at the center location, and (ii) holes of size $r_{\text{max}}h$ in which no points are present are identified and filled with a point at the center of the hole. In both cases, all physical properties are interpolated at a new point location. Here, $r_{\text{min}}$ and $r_{\text{max}}$ are fixed application-independent constants that determine the number of points in the support domain. For more details about the procedures of identification of holes, the interpolation, and the thinning/filling procedures in general, we refer to \cite{Drumm2008, SeiboldThesis, Suchde2019_MovingSurfaces}. The same distortion fixing algorithms are also used for spatial adaptivity of the discretization. For example, if large material property gradients are detected, $h$ can be reduced, which would automatically trigger the hole-filling algorithm to refine the point cloud in that region.

\subsection{Spatial derivatives}
\label{sec:GFDM}

Here, we compute numerical derivatives with the strong form meshfree generalized finite different method (GFDM) \cite{Basic2020, Fan2018, Liszka1980, Lu2016}. GFDM is a robust method and has been used for a variety of applications, especially in modeling fluid flow, see for example \cite{Jefferies2015, Michel2021, Moller2007, Tramecon2013}. The meshfree GFDM is a collocation approach, and as the name suggests, they generalize classical finite differences to arbitrarily distributed points \cite{Liszka1980}. The derivatives of a function $u$ are approximated by a linear combination of discrete function values, $u_i$ for each $i = 1, \dots, N$, of each neighboring point
\begin{equation}
	\label{Eq:GFDM_Definition}
	\partial^* u(\vec x_i)\approx \partial^*_i u = \sum_{j\in S_i}c_{ij}^* u_j,
\end{equation}
where ${}^*=x,y, z, \Delta$ represents the differential operator being approximated, $\partial^*$ represents the continuous $^*$-derivative, and $\partial^*_i$ represents the discrete derivative at point $i$. For each point $i$, the stencil coefficients $c_{ij}^*$ are found using a weighted least squares approach. The numerical differential operators can be found locally at each point by solving a small linear system, independently of the operators at the rest of the point cloud. Since the point cloud moves in time, the derivative stencils need to be recomputed for each point at every time step.

For a point $i$, consider Taylor expansions around $\vec{x}_i$ up to order $2$ terms for each neighbouring point $j\in S_i$
\begin{equation}
	\label{Eq:TaylorExp}
	e_{ij} + u(\vec{x}_j)=u(\vec{x}_i)+\nabla u \cdot (\vec{x}_j-\vec{x}_i) + \frac12 (\vec{x}_j-\vec{x}_i)^T D (\vec{x}_j-\vec{x}_i).
\end{equation}
The unknown coefficients of $\nabla u$ and $D$ are computed by a weighted least squares method by minimizing
\begin{equation}
	\label{Eq:minJ}
	\text{min } J_i = \sum _{j \in S_i} W_{ij}^2e_{ij}^2,
\end{equation}
where $W$ is a weighting function used to make sure that the points closer to the central point $i$ have a larger impact than the points farther away. We note that the square of the weighting function is used only for notational convenience. The weighting function is usually taken as a Gaussian distribution
\begin{equation}
	\label{Eq:WeigthingKernel}
	W_{ij} = \exp\!\left(-\alpha_W \frac{\|\vec{x}_j-\vec{x}_i\|^2}{h_i^2+h_j^2}\right),
\end{equation}
where $\alpha_W$ is a positive constant usually taken in the range of $(2,8)$. Note that the weighting function is only defined on the local support $S_i$ consisting of $n(i)$ points. For the sake of brevity, we present only the case of one spatial dimension. Eq.\,\eqref{Eq:TaylorExp} leads to the following system which is solved at each point $i=1,\dots, N$
\begin{equation}
	\label{Eq:GFDAsystem}
	\underbrace{\left(\begin{array}{c}
	e_{i1} \\
	\vdots \\
	e_{in}
	\end{array}\right)}_{\vec E_i} =
		\underbrace{ \left( 
		\begin{array}{ccc}
		1 & \delta x_{i1}     & \frac12\delta x_{i1}^2    \\
		\vdots & \vdots         & \vdots          \\
		1 & \delta x_{in}   & \frac12\delta x_{in}^2  \\
		\end{array}  \right) }_{M_i}
	\underbrace{\left(\begin{array}{c}
	(u_0)_i \\
	(u_x)_i \\
	(u_{xx})_i
	\end{array}\right)}_{\vec a_i} -
		\underbrace{\left(\begin{array}{c}
		u_1\\
		\vdots \\
		u_n
		\end{array}\right)}_{\vec b_i},
\end{equation}
where $\delta x_{ij} = x_j - x_i$. Or, in short form $\vec{E}_i = M_i\vec{a_i} - \vec{b}_i$. The minimization Eq.\,\eqref{Eq:minJ} can be rewritten as
\begin{subequations}
\begin{align}
	\text{min } J_i &= \vec E_i ^{\,T} W_i^2 \vec E_i\\
	&= (M_i\vec a_i-\vec b_i)^T W_i^2 (M_i\vec a_i - \vec b_i) 	\label{Eq:minJ2},
\end{align}
\end{subequations}
where $W_i$ is a diagonal matrix with entries $W_{i1},\dots,W_{in}$. A formal minimization leads to
\begin{equation}
	\label{Eq:FinalSolution_GFDM}
	\vec a_i = [(M_i^T W_i^2 M_i)^{-1} M_i^T W_i^2 ]\vec b_i.
\end{equation}
This leads to the differential operator stencils as in Eq.\,\eqref{Eq:GFDM_Definition}
\begin{subequations}
 \begin{align}
	(u_x)_i &= \sum_{j \in S_i} c_{ij}^x u_j,\label{Eq:ux}\\
 	(u_{xx})_i &= \sum_{j \in S_i} c_{ij}^{xx} u_j, \label{Eq:uxx}
\end{align}   
\end{subequations}
where $c_{ij}^x$ and $c_{ij}^{xx}$ represent the values in the second and third row respectively of the matrix $[(M_i^T W_i^2 M_i)^{-1}M_i^T W_i^2]$ in Eq.\,\eqref{Eq:FinalSolution_GFDM}. 

We note that similar GFDM approaches have already been used to simulate phase change processes \cite{Resendiz2018, Saucedo2019, Veltmaat2022}. They model solidification at low velocities, while the present work models a lot more turbulent phase change process in vaporization. Furthermore, \cite{Resendiz2018, Saucedo2019, Veltmaat2022} do not take the latent heat into account directly, as done in the presented model. 

\subsubsection{Polynomial Formulation}~\\
\label{sec:PolynomialFormulation}
We now rewrite the Taylor expansion-based derivation to a polynomial formulation that can be easily extended for our purpose. An alternative, but equivalent, way to obtain the stencil coefficients in  Eq.\,\eqref{Eq:ux} -- Eq.\,\eqref{Eq:uxx} is to ensure that the derivatives of monomials $m\in\mathcal{M}$ up to the order of accuracy desired are exactly reproduced, leading to the optimization problem
\begin{subequations}
\begin{align}
	\sum_{j\in S_i}c_{ij}^{*}m_j &= \partial^* m (\vec{x}_i),\qquad \forall m\in\mathcal{M},\label{Eq:Consistency}\\
	\text{min } J_i &= \sum_{j\in S_i} \left( \frac{c_{ij}^{*}}{W_{ij}} \right)^2. \label{Eq:BasicMin}
\end{align}
\end{subequations}
where ${}^* \in \{x, y, xx, \Delta, \dots \}$. 

In our earlier work \cite{Suchde2018_Thesis}, we have proven the equivalence of these formulations. An efficient method to compute the differential operator stencils in this formulation using a QR decomposition can also be found in \cite{Suchde2018_Thesis}.

\subsection{Enrichment}
\label{sec:Enrichment}

Due to the smooth shape functions and the strong form nature of the GFDM, representing discontinuities is not straightforward. Using the polynomial formulation of Section~\ref{sec:PolynomialFormulation}, we now introduce an enrichment mechanism that is needed to accurately capture the sharp changes in the material properties, which is especially necessary for handling jumps and discontinuities. Motivated by our earlier work for two-dimensional manifolds \cite{Suchde2019_StaticSurfaces}, we extend the polynomial space of the test functions to include test functions with discontinuities. 

We note that the differential operator for which discontinuities play the largest role is the $\nabla \cdot ( \alpha \nabla \phi )$ for some material property $\alpha$ and some primary variable $\phi$. For example, the term $\nabla \cdot ( \frac{1}{\rho} \nabla p )$ arises in the pressure Poisson equation,  and $\nabla \cdot ( \lambda_{\text{eff}} \nabla T )$ in the energy conservation equation. Similarly, the momentum conservation equation contains the divergence of the stress tensor which includes the term $\nabla \cdot \left( \eta_{\text{eff}} \nabla  \left( \left( \nabla \vec{v} \right) + \left( \nabla \vec{v} \right)^T  \right) \right)$. Each of these contains a diffusion operator of the form $\nabla \cdot ( \alpha \nabla \phi )$. Here, we introduce an enrichment of this numerical differential operator to handle jump discontinuities in material properties. 

Using the shorthand $D^{\alpha}\phi = \nabla\cdot(\alpha\nabla\phi)$, the diffusion operator is computed as
\begin{subequations}
 \begin{align}
	\sum_{j\in S_i}c_{ij}^{D^{\alpha}}m_j &= D^{\alpha} m (\vec{x}_i),\qquad \forall m\in\mathcal{M},\label{Eq:DEG_Consistency}\\
	\text{min } J_i &= \sum_{j\in S_i} \left( \frac{c_{ij}^{D^{\alpha}}}{W_{ij}} \right)^2, \label{Eq:DEG_Min}
\end{align}   
\end{subequations}
in the polynomial formulation. The space $\mathcal{M}$ of consistency conditions in Eq.\,\eqref{Eq:DEG_Consistency} is enhanced with discontinuous functions, in a manner similar to that done in \cite{Yoon2014, Yoon2014b} in the context of crack propagation. For $\vec{s} = \frac{\nabla \alpha}{\| \nabla \alpha \|}$ denoting the normalized gradient of the material property being considered, the extra conditions added are given by
\begin{subequations}
\begin{align}
	\sum_{j\in S_i}c_{ij}^{D^{\alpha}} \frac{1}{\alpha_j} &=  - \Delta \log \alpha,\\
	\sum_{j\in S_i}c_{ij}^{D^{\alpha}} \frac{\delta s_{ij}}{\alpha_j} &= - \frac{\partial}{\partial s} \log \alpha,\\
	\sum_{j\in S_i}c_{ij}^{D^{\alpha}} \frac{\left( \delta s_{ij} \right)^2}{\alpha_j} &= 2,
\end{align}
\end{subequations}
where $\delta s_{ij} = \vec{s} \cdot \delta \vec{x}_{ij}$, and $\frac{\partial}{\partial s} = \vec{s} \cdot \nabla$. 


%
%


\section{Numerical Scheme}
\label{sec:TimeInteg}

The discretization of all spatial derivatives is done as explained in Section~\ref{sec:GFDM}. The time integration of the constitutive equations in Section~\ref{sec:Equations} is done using a segregated approach which forms a Chorin-type projection method \cite{Brown2001, Chorin1968}. The integration begins with the Lagrangian movement of the point cloud, done with a second-order method~\cite{Suchde2018_PCM}. Then an intermediate velocity is computed by a first-order implicit discretization of the momentum equation. This is then projected to the space of the desired velocity divergence with the help of a pressure Poisson equation, followed by a pressure update. After this, a Crank-Nicolson time-integration scheme is used for the energy conservation followed by an implicit integration of the turbulence equations \cite{Michel2021}. For details on the entire time-integration scheme, we refer to our earlier work \cite{Drumm2008, Jefferies2015, Kuhnert2014, Michel2021}.

\subsection{Temperature smoothing}

The strong form nature of the derivative computation can lead to numerical instabilities while trying to capture jumps. Since the temperature field can have jump discontinuities due to the phase change process, we numerically smooth the temperature with a Gaussian smoothing kernel \cite{Suchde2019_MovingSurfaces}. The smoothed temperature field at a point $i$ is given by
\begin{equation}
	\widetilde{T}_i = \frac{ \sum_{j \in S_i} K_{ij} T_j}{ \sum_{j \in S_i} K_{ij}},
\end{equation}
where the smoothing kernel is given by
\begin{equation}
	K_{ij} = \exp\left( -2 \frac{\| \vec{x}_j - \vec{x}_i \|^2}{h_i^2 + h_j^2}  \right).
\end{equation}
All material properties in Section~\ref{sec:Material} dependent on temperature are taken to be dependent on $\widetilde{T}$ numerically. As a result, this smoothed temperature field automatically has a smoothing effect on, for example, the numerical density field. Empirically, we observe that this helps ensure numerical stability.

\subsection{Volume expansion}

An important challenge in this macroscopic phase change modeling is to accurately capture the volume expansion (or contraction) as a result of the phase change. If a numerical point is in the liquid state at a certain time step, and subsequently in the vapor state in the next time step, this could result in a large time gradient in the density, which could lead to numerical instabilities when using the strong form GFDM. To overcome this issue, we rewrite the mass conservation in logarithmic form
\begin{subequations}
\begin{align}
	 \nabla \cdot \vec{v}\, \vert^{(n+1)} &= - \frac{d}{dt} \left( \log ( \rho^{(n+1)} ) \right),\\
	 &\approx - \frac{1}{\Delta t} \left( \log ( \rho^{(n+1)}) - \log ( \rho^{(n)}) \right),
\end{align}
\end{subequations}
which is used as the desired divergence of velocity in the projection step and the pressure Poisson equation. 



%
%

\section{Comparison with interface methods}
\label{sec:Comparisons}

Verification of the present one-fluid phase change model used here can not be done using standard numerical benchmark cases used in interface capturing or tracking methods. This holds for the benchmarks used by both two-fluid methods, and one-fluid methods with explicit interface capturing or tracking, which we club together under the term \emph{interface methods}. There are several modeling choices made by these interface methods (see, for example, \cite{Das2015, Hardt2008, Rajkotwala2019, Tryggvason2015}) that do not have a one-to-one correspondence in the one-fluid model considered here. 

\begin{enumerate}
	\item Interface methods typically require boundary conditions at the interface. For the temperature field, this could be a Dirichlet boundary condition or a prescribed heat flux. Occasionally, velocity boundary conditions are also imposed. This is not possible in the present one-fluid model since there is no explicit boundary between the phases.
	\item The interface methods often use verification cases that distinguish between liquid and vapor phases at saturation temperature. Whereas, in the present model, we only have a fluid at saturation temperature that cannot be distinguished between liquid and vapor. Furthermore, many verification tests in the literature assume an initial condition of a ``super-saturated" liquid in which a liquid phase is present above the boiling point. This, once again, is not directly possible in the present approach. 
    We note here that since the aim of this work is to model phase change in highly dynamic fluid flow, the topic of modeling metastable super-heated liquid states is not relevant.
	\item A significant advantage of the present one-fluid model is that it can be used to simulate scenarios where initially only a liquid phase is present without any vapor phase. To achieve this, interface methods assume a very small vapor phase at the initial state of the simulation. However, in applications where the boiling process occurs at a fluid free surface (see the jet impinging on a hot plate test case in Section~\ref{sec:JetPlate}), assuming the presence of a vapor phase near an evolving free surface is not feasible. 
\end{enumerate}



\section{Numerical Simulations}
\label{sec:Simulations}

All the numerical methods mentioned above have been incorporated into the in-house software suite MESHFREE~\cite{MESHFREE}. 
The verification of the basic numerical schemes and the validation of the underlying numerical framework for fluid flow can be found in our earlier work \cite{Drumm2008, SeiboldThesis, Suchde2018_INSE}. In all the test cases considered below, radiative heat transfer is neglected. We further note that all simulations are full three-dimensional simulations. However, pseudo-two-dimensional clips of the results are shown for ease of visualization. 

\subsection{Vapor film growth around a heated sphere}

\begin{figure}
    \centering
    \includegraphics[width=0.32\textwidth]{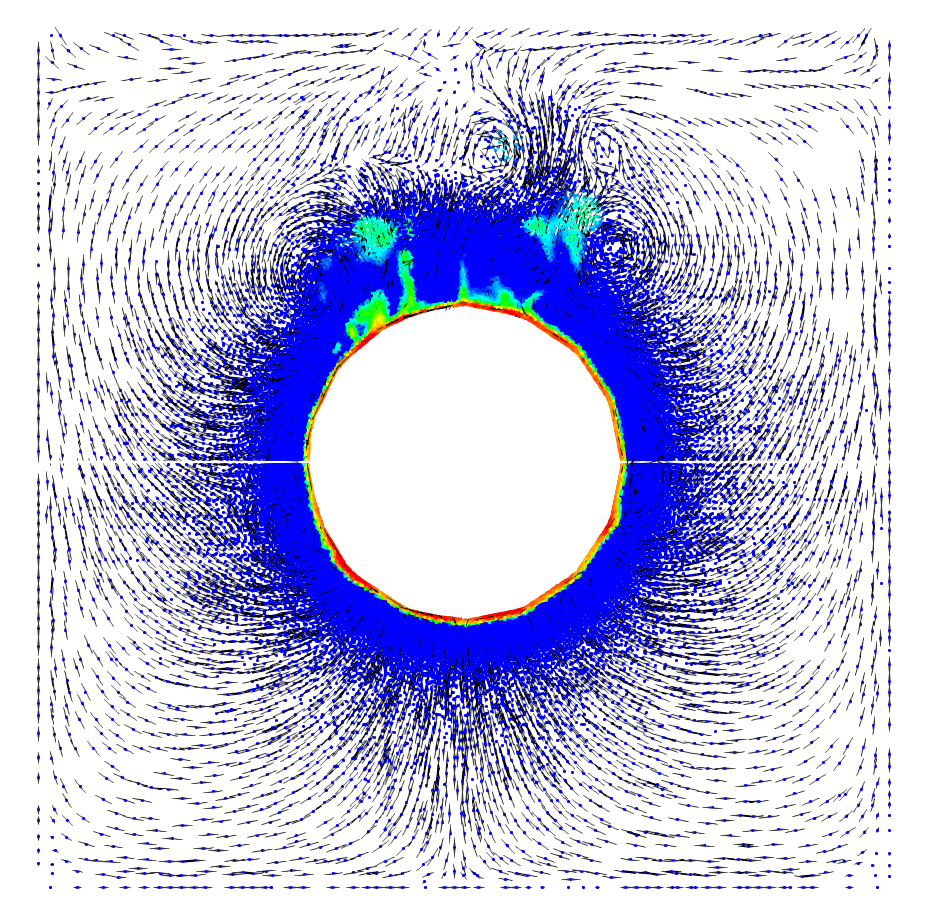}
    \includegraphics[width=0.32\textwidth]{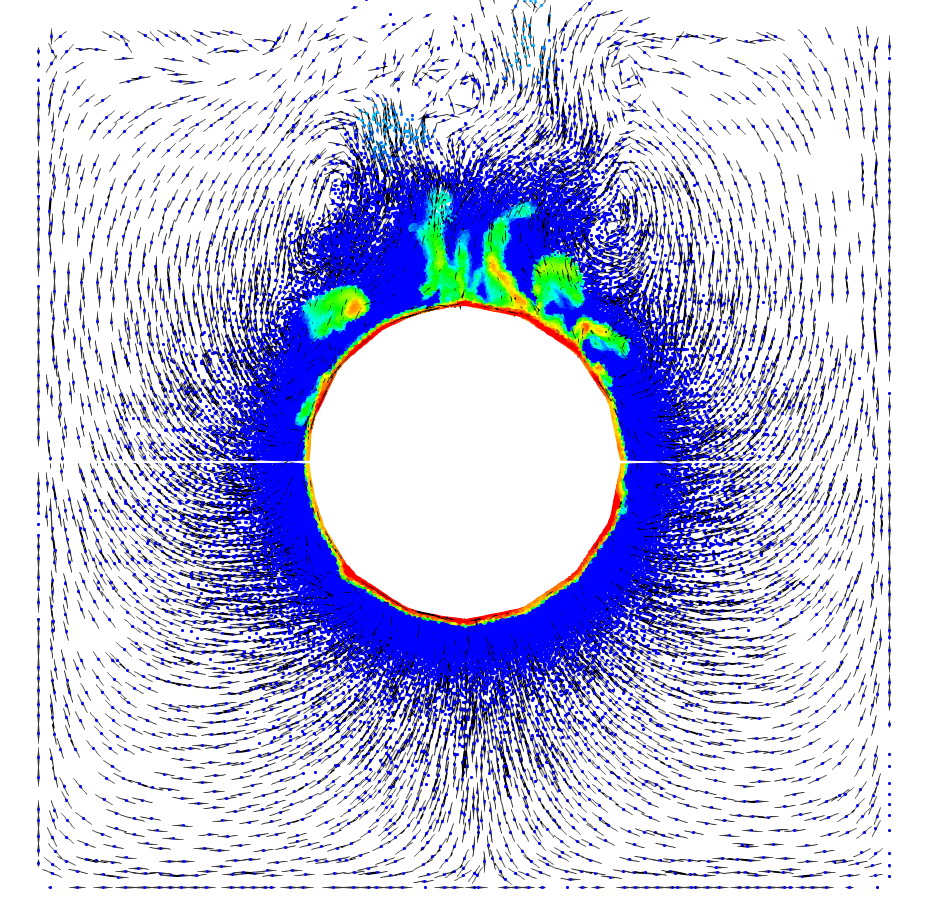} 
    \includegraphics[width=0.32\textwidth]{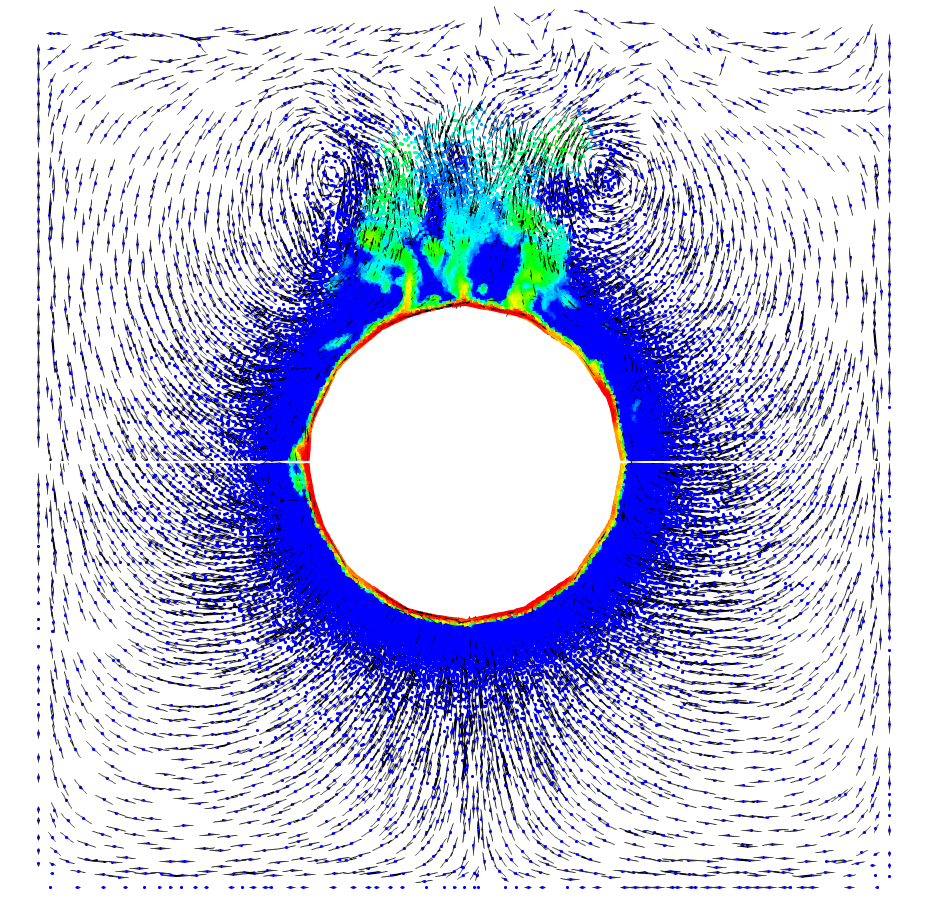}  
    \caption{Vaporization around a heated sphere: Formation of unsteady vapor films, bubble growth and rising under gravity. The figures show a clip of the three-dimensional domain at increasing times. The color indicates the value of $x^{\text{vapor}}$, with the red points representing higher values and the blue points representing lower ones. The arrows show the direction of the velocity. A zoomed-in version of the bubble formation is shown in Figure~\ref{Fig:Sphere_BubbleMovement}.}
    \label{Fig:Sphere_BubbleFormation}%
\end{figure}

We start by considering the growth of a vapor film around a heated solid sphere, shown in Figure~\ref{Fig:Sphere_BubbleFormation}, as also considered by \cite{Das2015}. A heated sphere is immersed in the middle of a fluid in a tank with an open top. The fluid tank has a square base area with a side length of \qty{4}{\centi\meter}. The tank is filled with a resting liquid of initial temperature \qty{89}{\degreeCelsius} up to a height of \qty{4}{\centi\meter}. A solid sphere with a diameter of \qty{1.5}{\centi\meter} is submerged in the middle of the tank. 

\emph{Simulation parameters:} 
The liquid and vapor densities are taken to be constant at $\rho^{\text{liquid}} = \qty{1000}{\kilogram\per\cubic\meter}$ and $\rho^{\text{vapor}} = \qty{10}{\kilogram\per\cubic\meter}$ respectively. The phase change takes place within the vaporization interval of $\Delta T^* = \qty{5}{\degreeCelsius}$ around the boiling point of $T_p^* = \qty{100}{\degreeCelsius}$, while neglecting pressure dependence of the boiling point. The specific heat capacities are $c_{v}^{\text{liquid}} = \qty{4184}{\joule\per\kilogram\per\kelvin}$ and $c_{v}^{\text{vapor}} = \qty{2030}{\joule\per\kilogram\per\kelvin}$ for the liquid and vapor, respectively. The latent heat of vaporization is $\Delta H_p^{\text{vap}} = \qty{2501}{\kilo\joule\per\kilogram\per\kelvin}$. The heat conductivities are $\lambda^{\text{liquid}} = \qty{0.6}{\watt\per\meter\per\kelvin}$ and $\lambda^{\text{vapor}} = \qty{0.05}{\watt\per\meter\per\kelvin}$ for the liquid and vapor respectively. The liquid has a surface tension of $\sigma = \qty{0.07286}{\newton\per\meter}$ and a viscosity of $ \eta^{\text{liquid}} = \qty{0.001}{\pascal\second}$, while the vapor has a viscosity of $ \eta^{\text{vapor}} = \qty{0.00013}{\pascal\second}$, with no surface tension. The domain is resolved with a coarse resolution of $h_0$ and a fine resolution of $h_{\min} = 0.1 h_0$ near the sphere surface. The resolution of the point cloud is gradually increased radially at a rate of $dh / dr = 0.15$. At the walls and the free surface at the top, we apply a zero heat flux boundary condition. The heat transfer between the solid sphere and the fluid is governed by a heat transfer coefficient of $U_s = \qty{12500}{\watt\per\square\meter\per\kelvin}$. The temperature on the sphere surface $T_S$ is growing linearly in time from $T_S(\qty{0}{\second}) = T^* + \Delta T^*$ to $T_S(\qty{6}{\second}) =  \qty{400}{\degreeCelsius}$.

\begin{figure}
    \centering
    \includegraphics[width=0.32\textwidth]{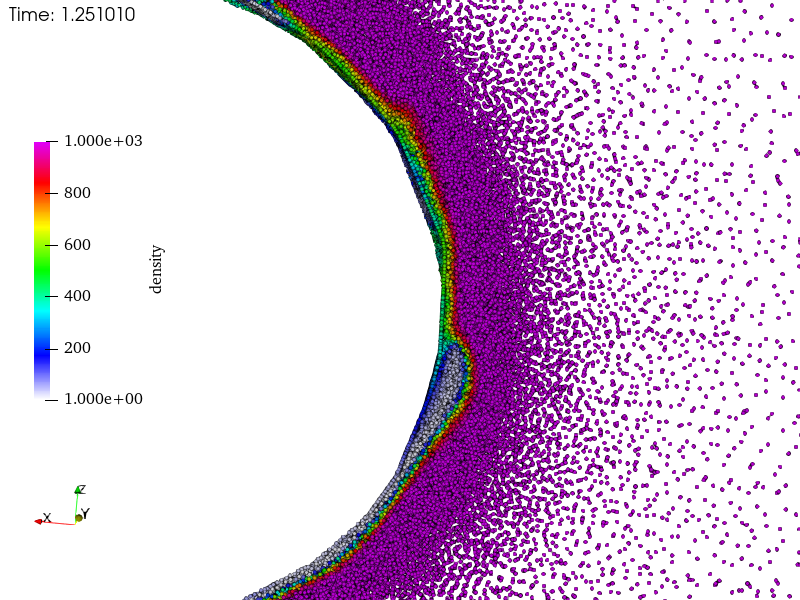}
    \includegraphics[width=0.32\textwidth]{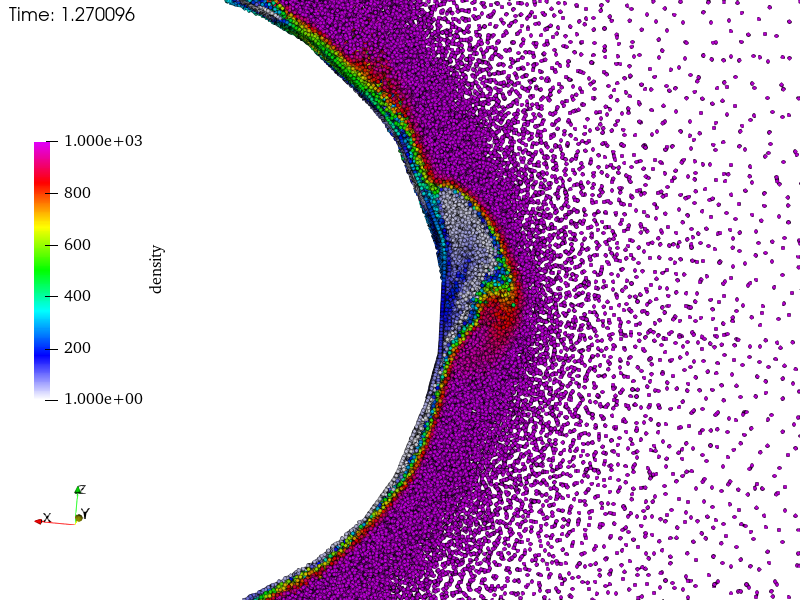}    
    \includegraphics[width=0.32\textwidth]{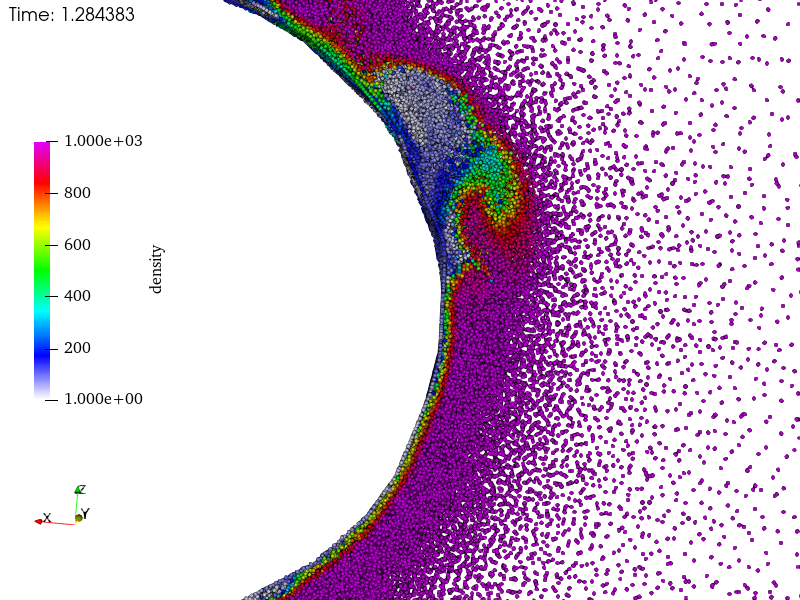}  
    \caption{Vaporization around a heated sphere: Bubble growth around an unsteady vapor film layer. The color shows the density. A zoomed-out version of this can be found in Figure~\ref{Fig:Sphere_BubbleFormation}.}
    \label{Fig:Sphere_BubbleMovement}%
\end{figure}

\begin{figure}
    \centering
    \includegraphics[width=0.32\textwidth]{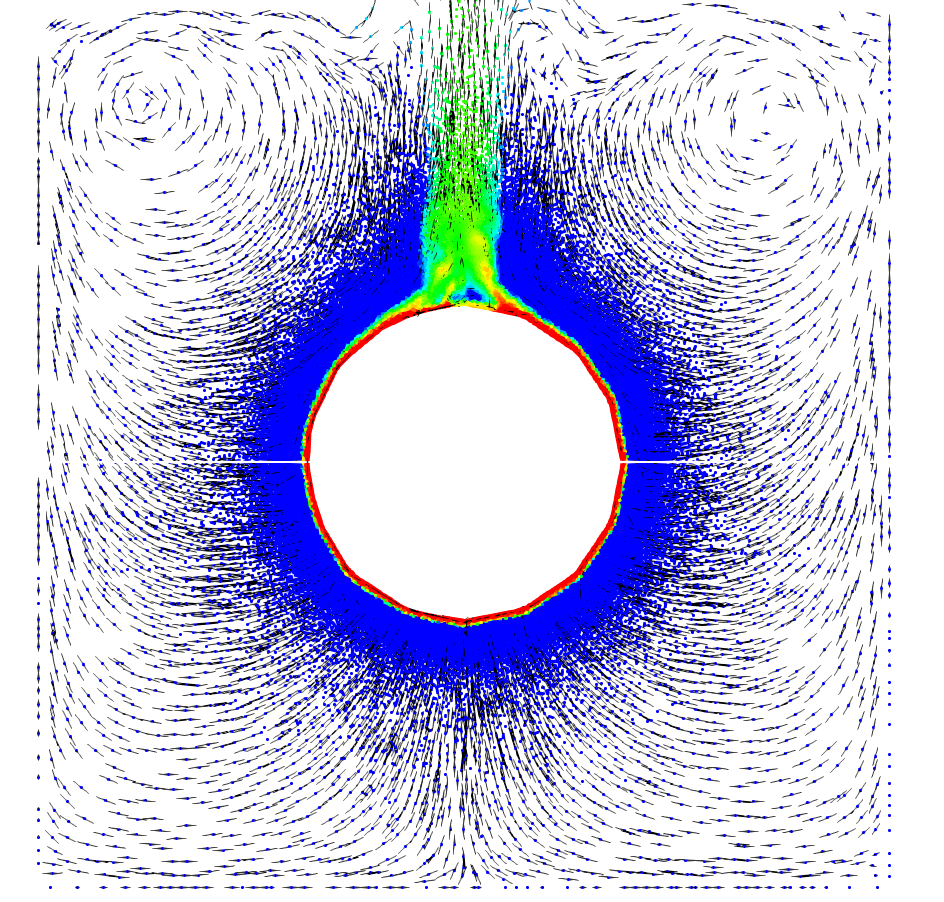}
    \caption{Vaporization around a heated sphere: Formation of thin, but steady overheated vapor layer and rising flow under gravity. This is the Leidenfrost-state, where $x^{\text{vapor}}$ tends to 1 around the sphere surface.}
    \label{Fig:Sphere_LeidenfrostState}%
\end{figure}

\begin{figure}
  \centering
  \includegraphics[width=0.7\textwidth]{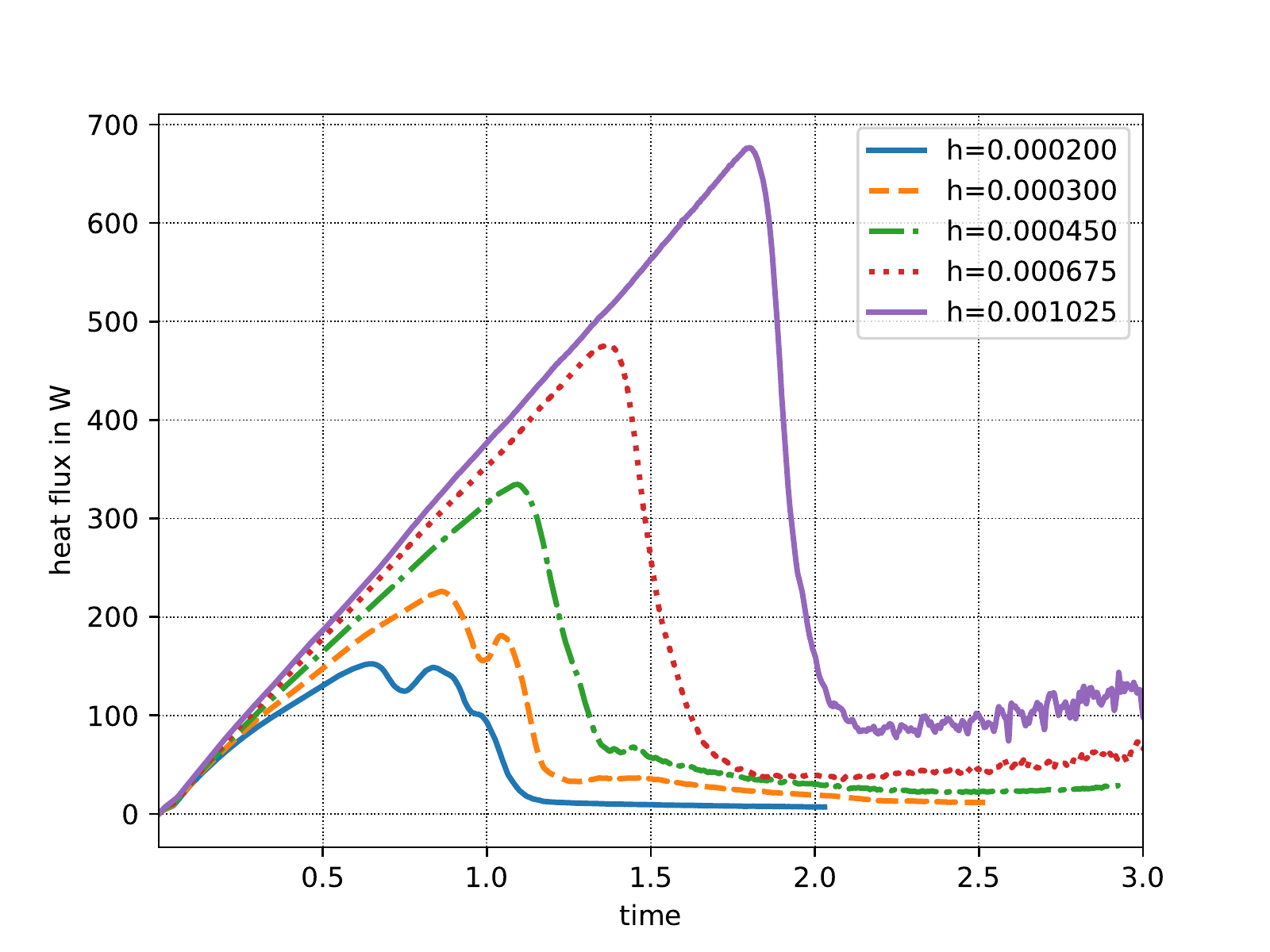}    
	\caption{Vaporization around a heated sphere: Heat flux across the sphere surface to the liquid.}
  \label{Fig:Sphere_HeatFlux}%
\end{figure}

A thin vaporization area forms initially with relatively small values of $x^{\text{vapor}}$. The lower density of the vapor causes it to rise under gravity. As temperature $T_S$ increases, the values of $x^{\text{vapor}}$ rise, which results in the formation of vapor bubbles around the film, which leave the film as they rise and break at the free surface of the fluid. The vapor bubble formation and breaking are shown in Figure~\ref{Fig:Sphere_BubbleFormation}, with Figure~\ref{Fig:Sphere_BubbleMovement} showing a detailed sketch of a particular bubble formation. With rising $T_S$, the fluid in the vicinity of the sphere tends to $x^{\text{vapor}}=1$, which provides an isolation layer around the sphere, illustrated in Figure~\ref{Fig:Sphere_LeidenfrostState}. The lower heat conductivity of the vapor phase results in a rapid drop in the heat flux from the hot sphere to the fluid. These effects are also observed in experiments and are often referred to as the Leidenfrost-effect which prevents the liquid beyond the vapor film from boiling. This effect is quantified in Figure~\ref{Fig:Sphere_HeatFlux} which shows the integrated heat transfer over the surface of the sphere for different resolutions. As the temperature rises, the heat transfer is rising for all resolutions. However, for each resolution, we observe the aforementioned rapid Leidenfrost-drop of the heat transfer at different temperatures. For finer resolutions, the temperatures of the drop come closer to each other, however, it seems that the numerical resolution still does not meet the thickness of the real vapor layer as it would arise in reality. The temperature at which the Leidenfrost-effect occurs is dominated by the choice of $U_s$, so experiments around the Leidenfrost state could, in the future, be used for calibrating the heat transfer coefficient.
%
%
%

\begin{figure}
  \centering
  \includegraphics[width=0.4\textwidth]{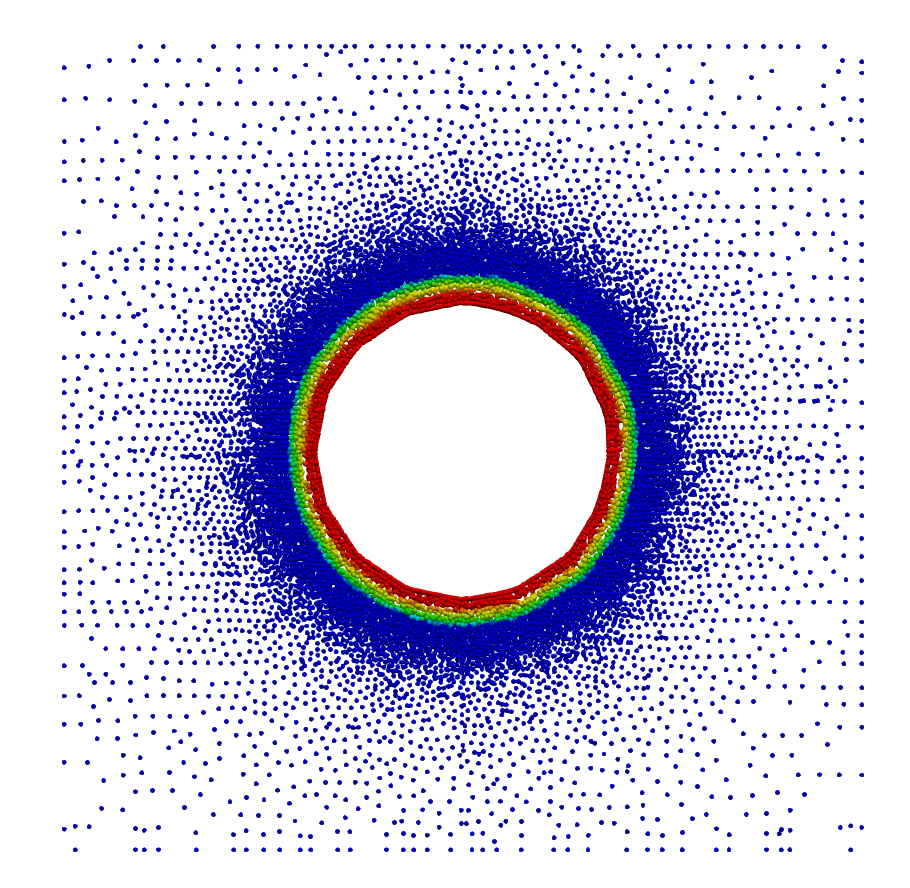}    
	\caption{Vaporization around a heated sphere without gravity: Formation of a uniform thickness vapor film, which grows slowly in time. Only a clip of the three-dimensional domain is shown. The color shows the value of $x^{\text{vapor}}$. }
  \label{Fig:Sphere_NoGravity}%
\end{figure}

As a verification, we also run this same test case in the absence of gravity. As also observed in \cite{Das2015}, without gravity, the vapor film formed is of a uniform thickness and is symmetric, see Figure~\ref{Fig:Sphere_NoGravity}. Unlike the case with gravity, the vapor film formed here is stable throughout the simulation without the formation of bubbles, even at lower temperatures of the sphere. 
%






\subsection{Cooking pot}
\label{Sec:CookingPot}





\begin{figure}
  \centering
  \includegraphics[width=0.32\textwidth]{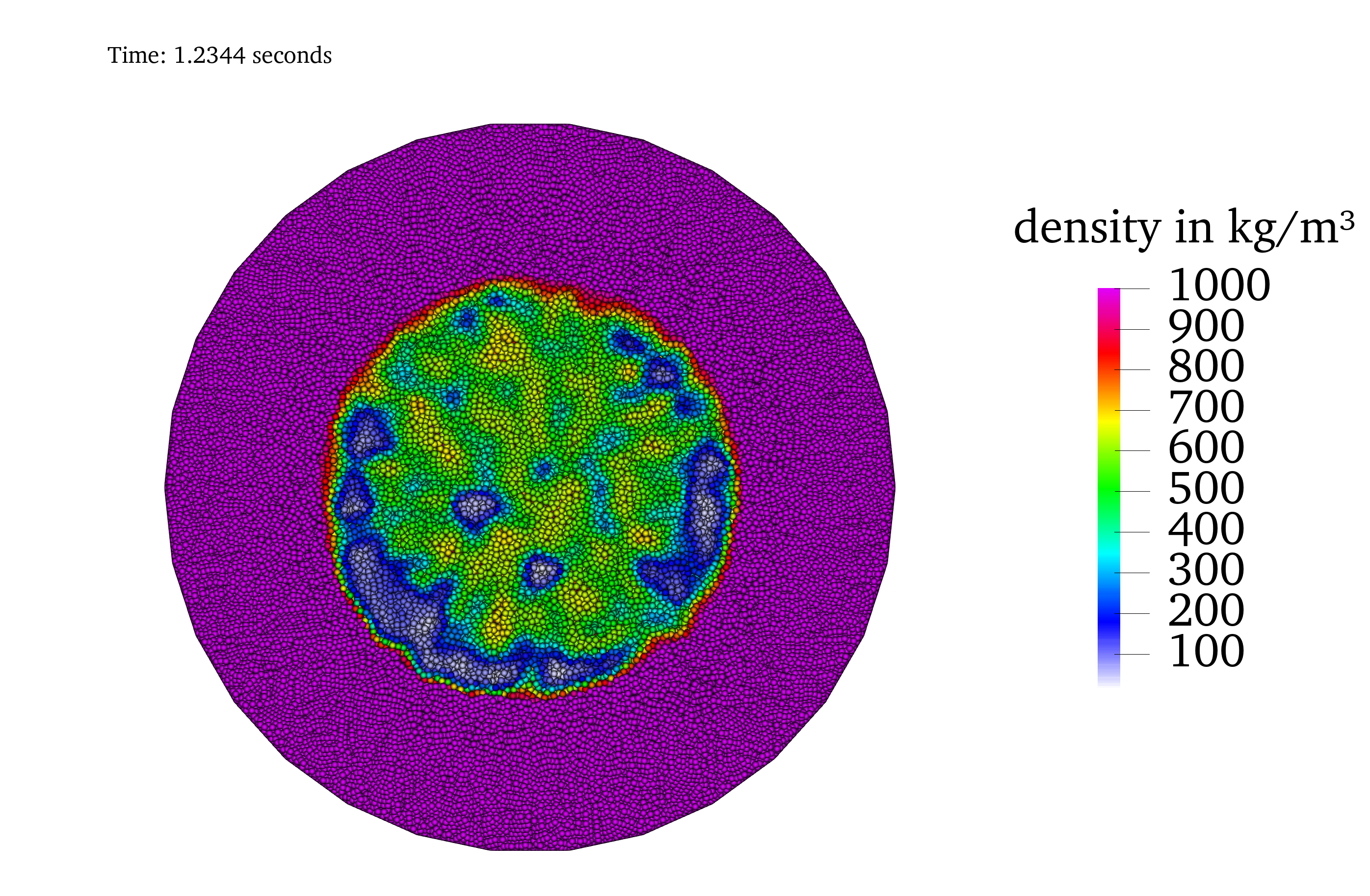}    
  \includegraphics[width=0.32\textwidth]{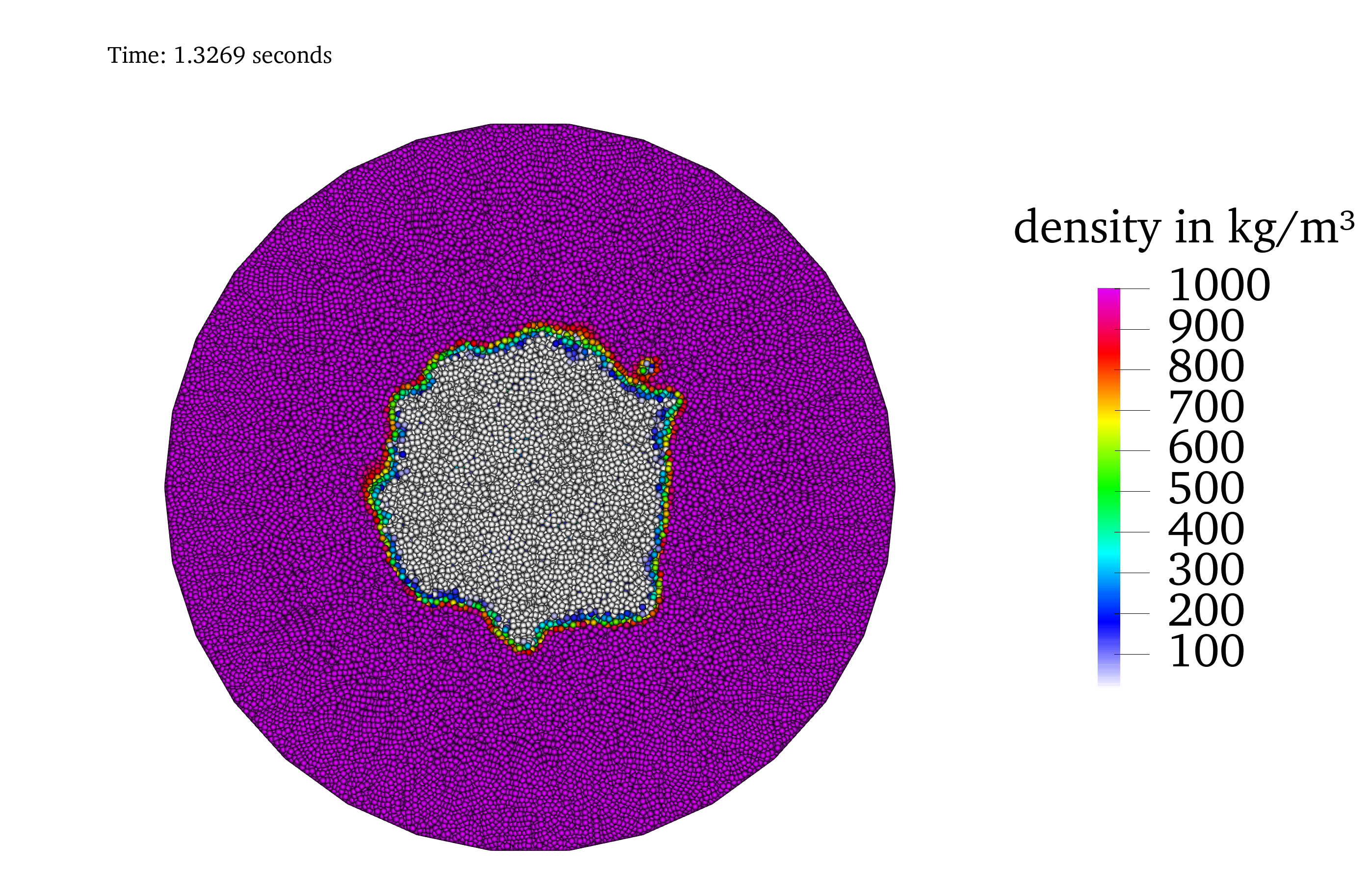}
  \includegraphics[width=0.32\textwidth]{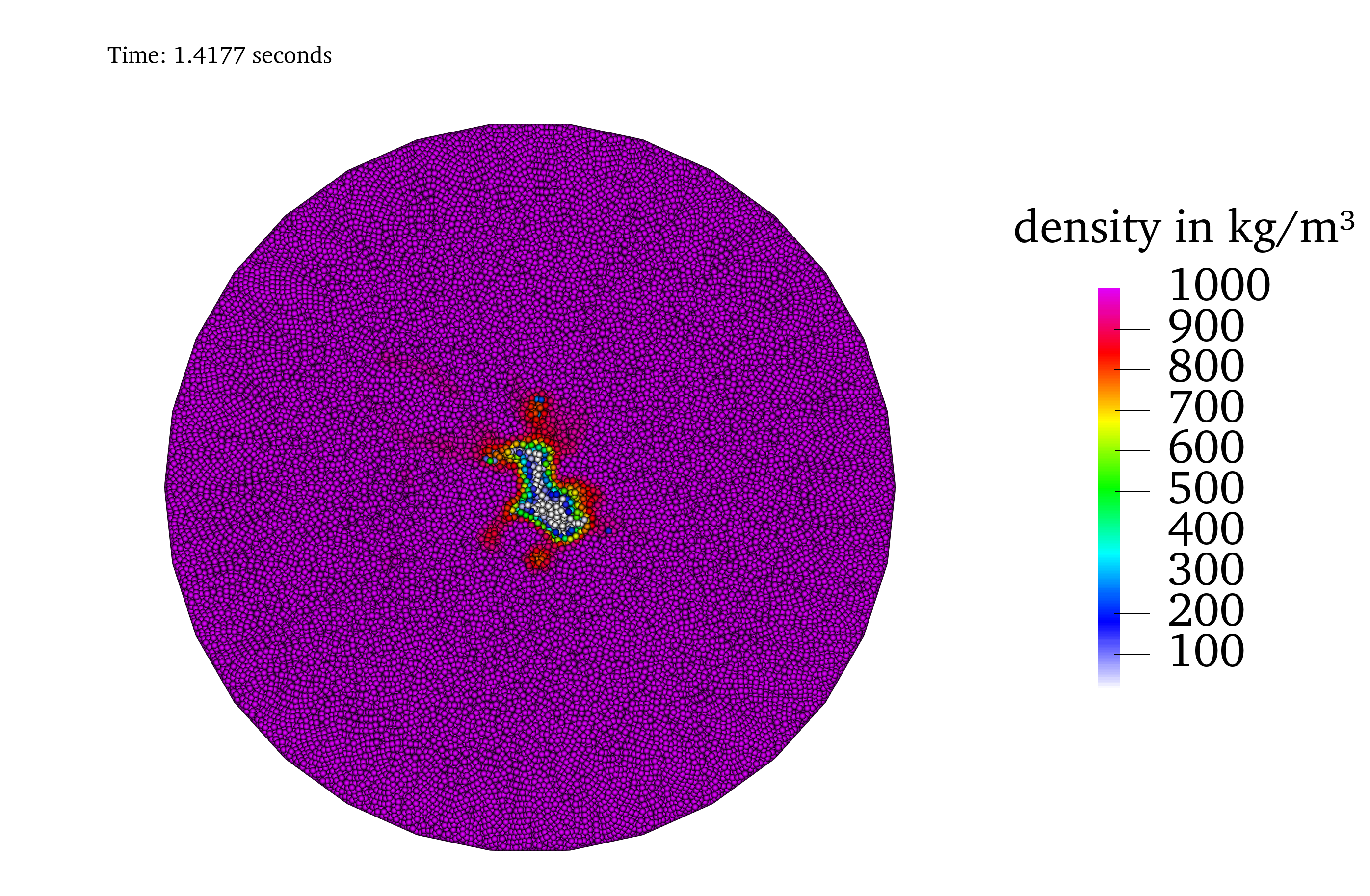}
  
  \includegraphics[width=0.32\textwidth]{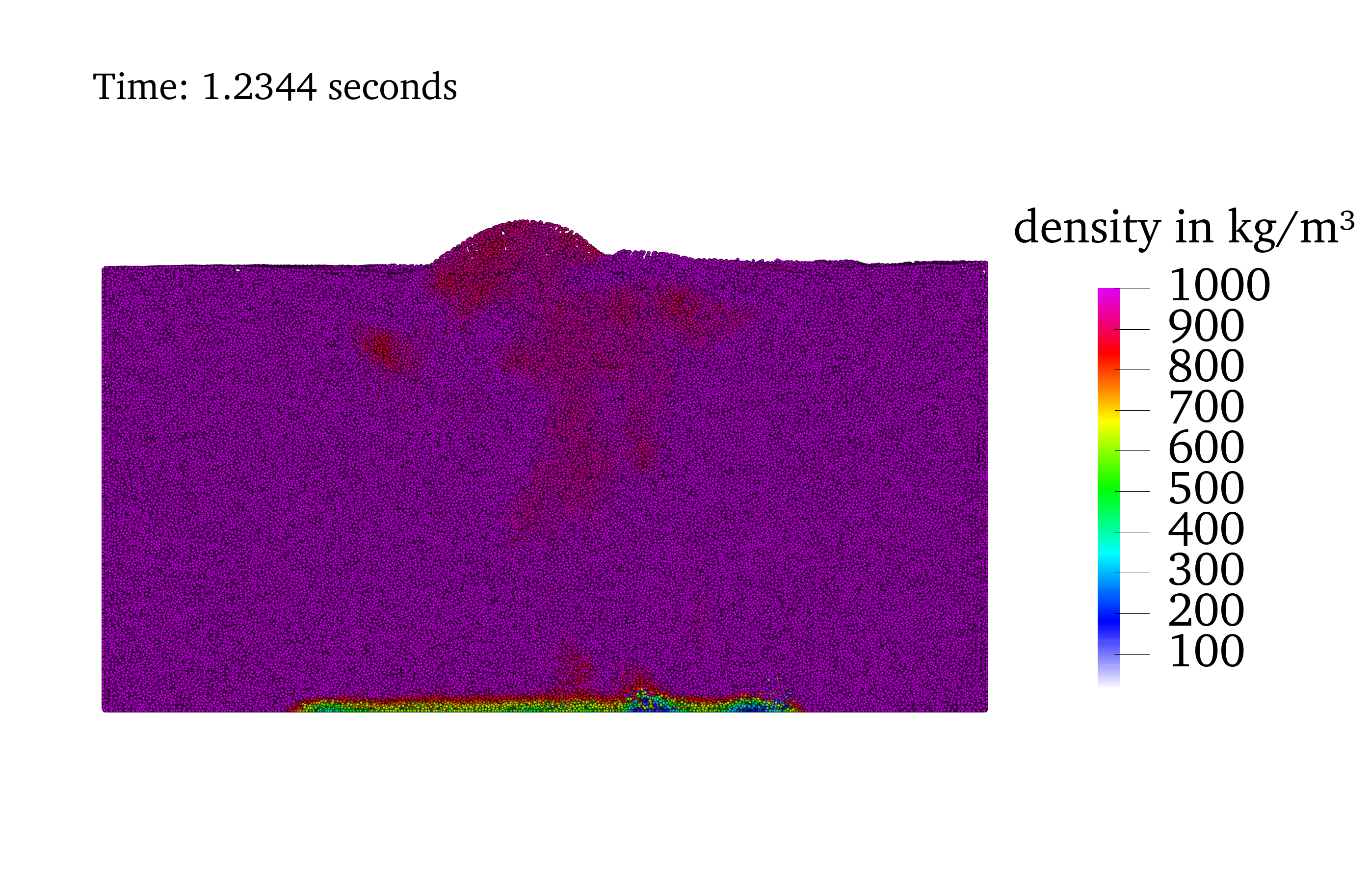}    
  \includegraphics[width=0.32\textwidth]{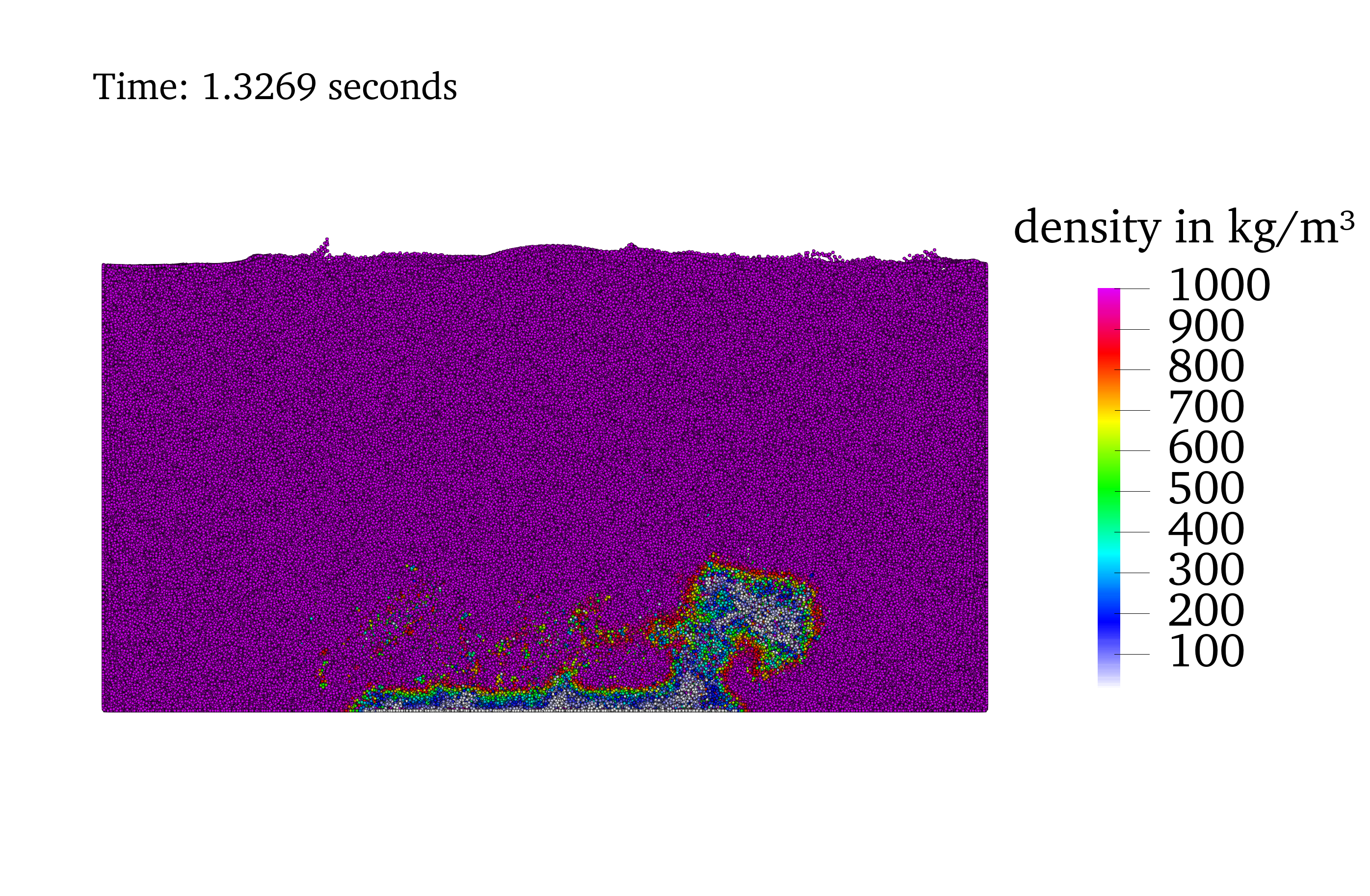}
  \includegraphics[width=0.32\textwidth]{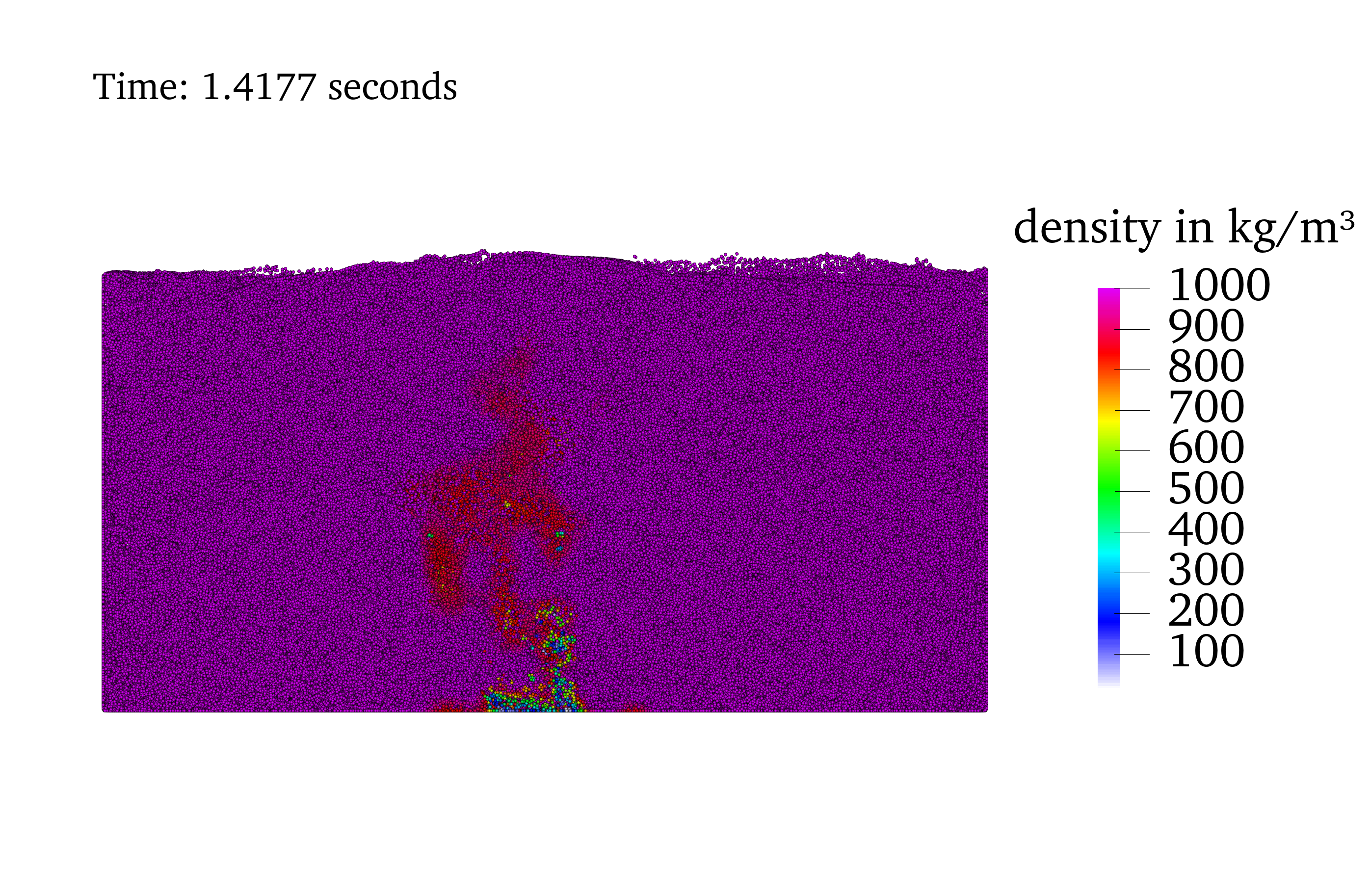}

	\caption{Boiling in a cooking pot: Snapshots of instantaneous density distribution at 1.2444 (left side), 1.3269 (middle), and 1.4177 seconds (right side) in the cooking bowl. (Top: bottom surface view, bottom: side view).}
  \label{Fig:HotPot_DensityDistribution}%
\end{figure}

We now simulate nucleate boiling in a cooking pot of diameter \qty{0.1}{\meter} filled with a liquid of initial temperature of \qty{85}{\degreeCelsius} to a height of $H = \qty{0.15}{\meter}$. An inner circle with diameter \qty{0.05}{\meter} in the middle of the bottom surface of the pot is heated by linear ramping to $T = \qty{400}{\degreeCelsius}$ in \qty{1}{\second}. The heat transfer coefficient at the hot wall is $U_s = \qty{12500}{\watt\per\square\meter\per\kelvin}$. On other surfaces, a zero heat flux boundary condition is imposed. The material properties of the liquid are as described in the previous test case. We simulate this case with a constant smoothing length over the liquid domain. Three simulations of $h = \qty{0.002}{\meter}, \qty{0.004}{\meter}$, and $\qty{0.008}{\meter}$ respectively are considered. The resulting evolution of the boiling in the cooking pot is shown in Figure~\ref{Fig:HotPot_DensityDistribution}.

\begin{figure}
  \centering
  \includegraphics[width=0.7\textwidth]{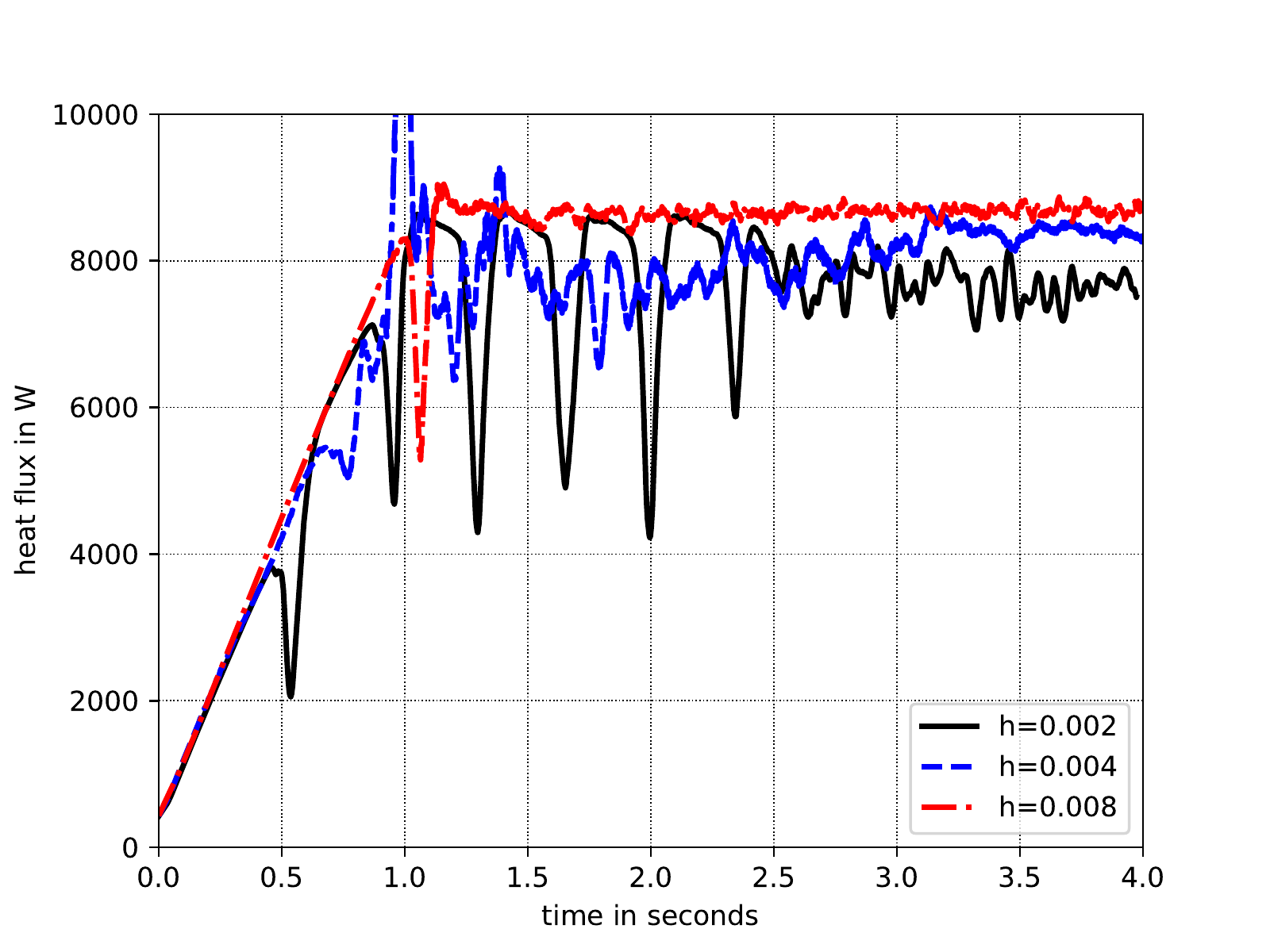}    
	\caption{Convergence study for heat flux in a cooking pot.}
  \label{Fig:HotPot_HeatFlux}%
\end{figure}

To quantify the results, we study the energy flux as a result of the fixed heat transfer coefficient. For all resolutions considered, the heat transfer is rising up to approximately $\qty{9000}{\watt}$ within the first second and then remains roughly constant. We observe temporary breakdowns of the heat transfer, especially for finer resolutions, see Figure~\ref{Fig:HotPot_HeatFlux}. They might be explained by temporary vapor films covering the heating zone at the bottom and thus acting as an insulating barrier, see Figure~\ref{Fig:HotPot_DensityDistribution}. The vapor film is metastable (Rayleigh-Taylor-like behavior) and quickly disappears. With progress in time, the breakdowns disappear even for the simulation with the finest resolution, which is due to the increasingly chaotic nature of the flow.

\begin{figure}
  \centering
  \includegraphics[width=0.49\textwidth]{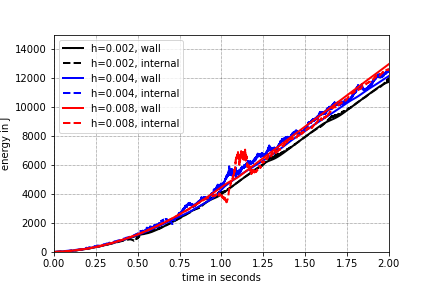}
  \includegraphics[width=0.49\textwidth]{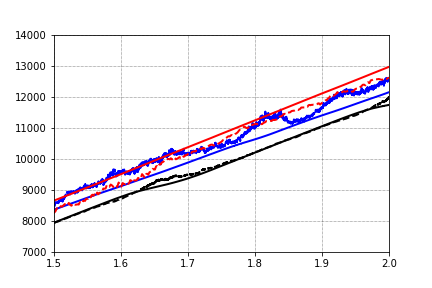}    
	\caption{Energy balance in the cooking pot. The figure on the right shows a zoomed in part of the figure on the left.}
  \label{Fig:HotPot_EnergyBalance}%
\end{figure}

The energy balance shown in Figure~\ref{Fig:HotPot_EnergyBalance} compares the internal heat of the fluid to the energy that enters the fluid through the walls. This energy conservation is important to verify in meshfree simulations since most meshfree methods often have issues with global conservative behavior \cite{Suchde2017}. The figure shows that changes in internal energy and the energy transferred through the walls are almost the same for finer resolutions. The measured internal energy contains fluctuations, which come from the measurement of internal energy as $E_{\text{internal}}=\int_V \rho h dV \approx \sum_i h_i \rho_i V_i $, where $V_i$ is an approximation of the volume taken up by a numerical point, which inherently has fluctuations. 

\subsection{Fluid jet on a hot plate}
\label{sec:JetPlate}

\begin{figure}
  \centering
  \includegraphics[width=0.4\textwidth]{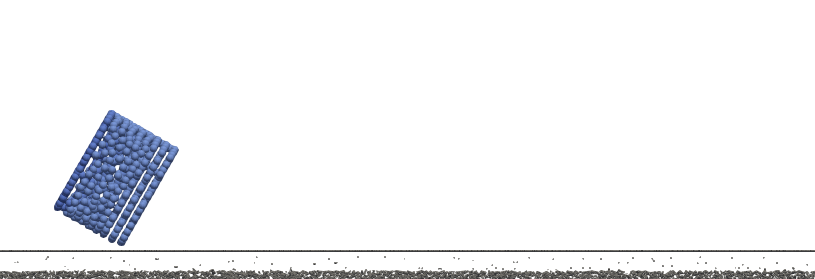}
  \includegraphics[width=0.4\textwidth]{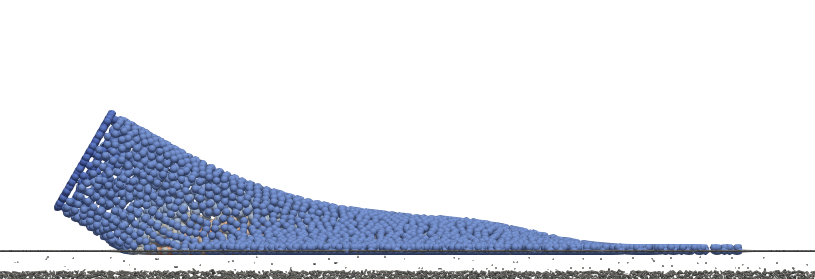} \\   
  \includegraphics[width=0.4\textwidth]{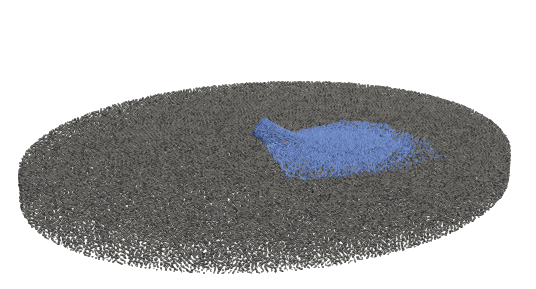}
  \includegraphics[width=0.4\textwidth]{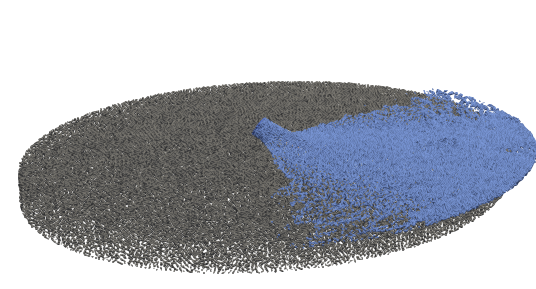}  
	\caption{Fluid jet impinging on a hot plate: Visualisation of the scenario being simulated \emph{without} the consideration of the phase change model. Side view~(top row), and top view~(bottom row) at different times. The solid plate is shown in gray, while the fluid (purely liquid) is shown in blue.}
  \label{Fig:JetImpinge}%
\end{figure}

%
%
As our third test case, we consider an example where boiling occurs at the free surface of the fluid, and compare simulations with experimental results. This example is motivated by the boiling of metalworking fluid introduced in Section~\ref{sec:Introduction}. The solid phase is a large circular plate made of Inconel 718 of diameter \qty{0.14}{\meter} and thickness of \qty{0.005}{\meter}, which is also discretized with a meshfree method, considering only energy conservation with zero velocity. A fluid jet is simulated with an impact angle of \qty{60}{\degree} on the solid plate. The inlet is considered very close to the plate (at a distance of \qty{0.005}{\meter} above) to save simulation time. The jet inlet has a diameter \qty{0.003}{\meter} and velocity \qty{7.0}{\meter\per\second}. 
The fluid impinging on a hot metal plate at an angle is very similar to typical conditions for simulating cooling jets in metal cutting applications \cite{Uhlmann2021b}.
These parameters are selected to match the experimental setup described by \cite{Nabbout2022}, except for the distance between the nozzle exit and the plate. The nozzle exit is placed closer to the plate to reduce the number of points in the simulation and thus, save computational costs. However, since all other parameters match the experimental setup, we expect no significant change in the global jet and especially the vaporization zones and cooling areas. 
For a visualization of this scenario being simulated, Figure~\ref{Fig:JetImpinge} shows the situation without the consideration of the vaporization model. 
%


\emph{Simulation parameters:} The vaporization interval is $\Delta T_p= \qty{5}{\degreeCelsius}$ around the boiling point at $T_p=\qty{100}{\degreeCelsius}$. The  latent heat of vaporization is $\Delta H_{p,vap} = \qty{2501}{\kilo\joule\per\kilogram\per\kelvin}$. Initially, the temperature of the plate is $T_0 = \qty{400}{\degreeCelsius}$. At the top face between the plate and the fluid, the heat transfer coefficient is $U_s = \qty{50000}{\watt\per\square\meter\per\kelvin}$. The ambient air is taken at $T = \qty{20}{\degreeCelsius}$. On all other walls of the solid, and at the inflow of the fluid phase we apply a zero heat flux boundary condition. A constant smoothing length is set at $h_l = \qty{0.00075}{\meter}$ for the liquid/vapor and at $h_s = \qty{0.0009}{\meter}$ for the solid.
The material properties are described in Table~\ref{Tab:MaterialProberties}. In contrast to the experiment, the vapor density is much larger. This is to avoid abrupt changes in the simulation. Hence, the vapor layer will be thinner. However, we do not expect a generally altered distribution and flow of the liquid. For this, we presume that the time scales of the expansion due to vaporization is much smaller than the time scales of convection. For the heat transfer by heat conduction, this means a different extent of the vapor layer. To counteract this effect we apply a decreased value of the heat conductivity, based on the ratio of simulated vapor density to real vapor density (\qty{0.50489}{\kilogram\per\cubic\meter}). 

\begin{table}
  \centering
  \begin{tabular}{cccccc}
	\hline
	\hline
	Phase & Density     & Surface     & Dynamic   & Specific           & Heat              \\
	      &       & tension     & viscosity & heat capacity      & conductivity      \\
	      & \unit{\kilogram\per\cubic\meter} & \unit{\milli\newton\per\meter} & \unit{\milli\pascal\second} & \unit{\joule\per\kilogram\per\kelvin} & \unit{\watt\per\meter\per\kelvin} \\
	\hline
	Liquid & 998  & 72.86 & 1    & 4185 & 0.59801 	 \\
	Vapor  & 100  & -     & 0.13 & 2059 & 0.00012405 \\
	Solid  & 8118 & -     & -    &  479 & 13.0       \\
	\hline
  \end{tabular}
  \caption{Material properties for the fluid jet on a hot plate test case.} \label{Tab:MaterialProberties}
\end{table}



\begin{figure}
  \centering
  \includegraphics[width=0.49\textwidth]{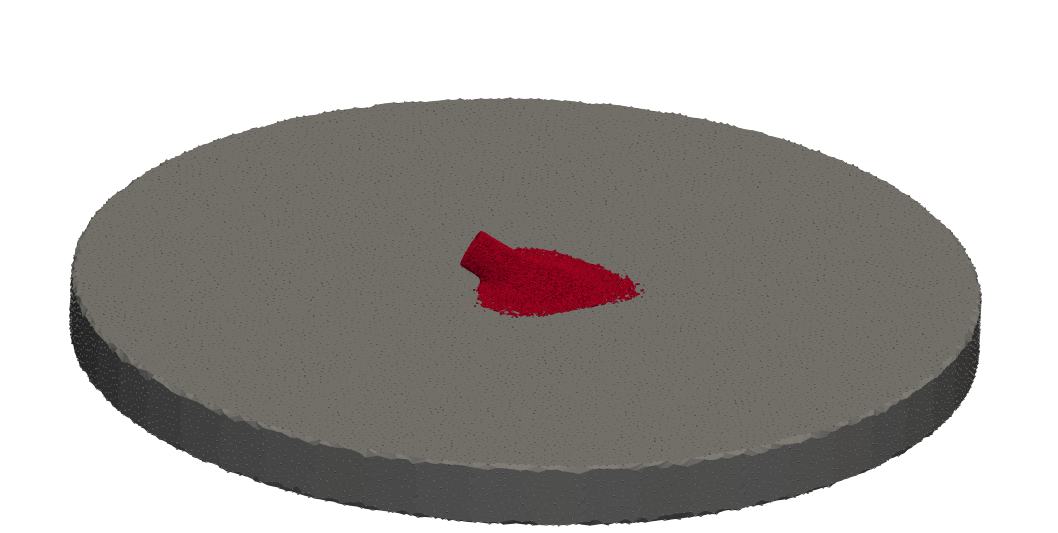}
  \includegraphics[width=0.49\textwidth]{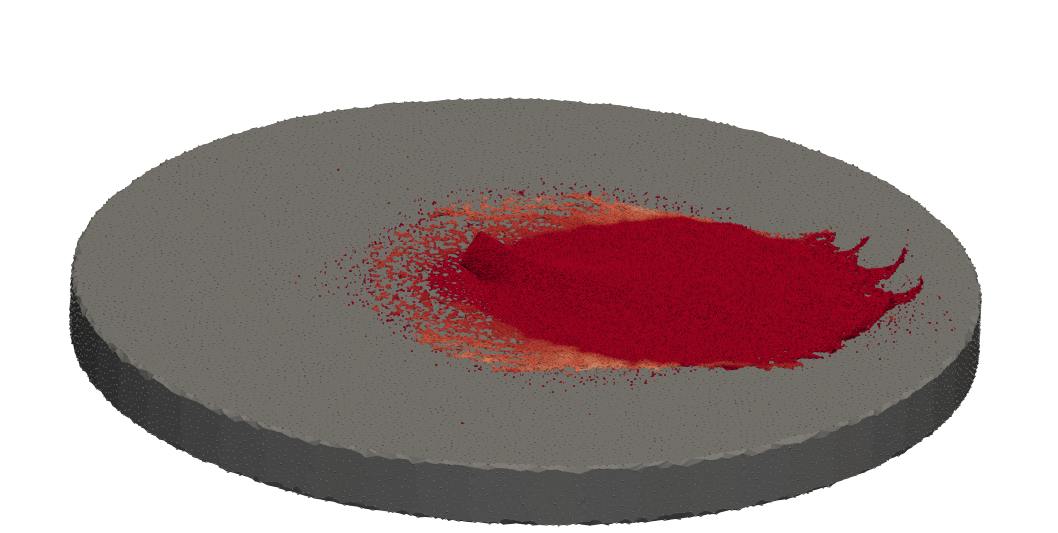}\\  
  \includegraphics[width=0.49\textwidth]{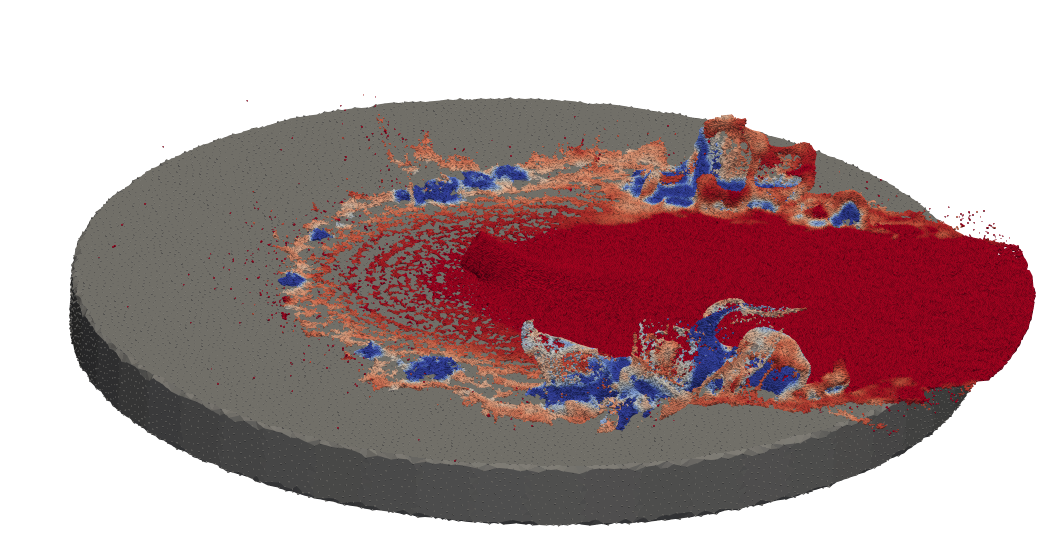}
  \includegraphics[width=0.49\textwidth]{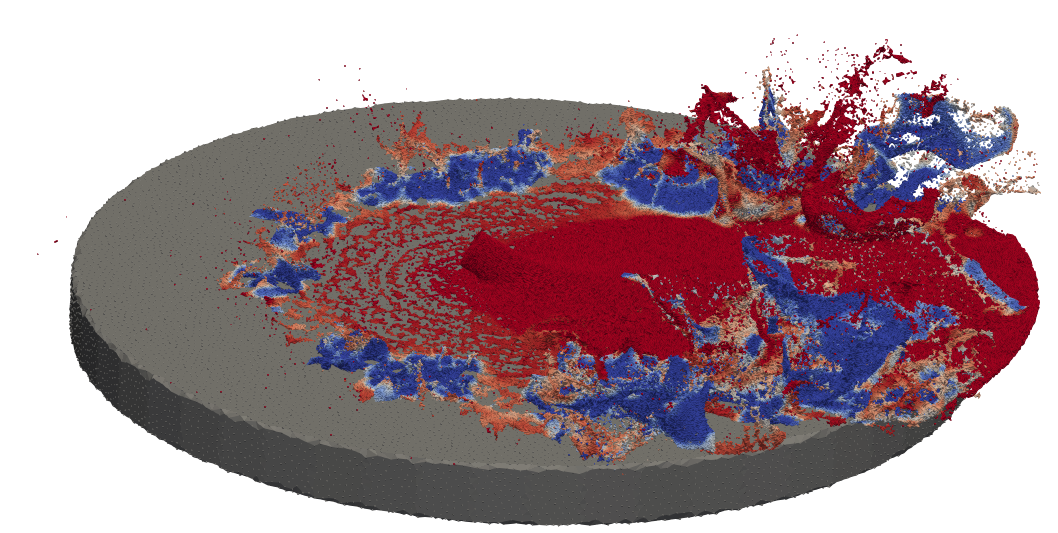}\\      
  \includegraphics[width=0.49\textwidth]{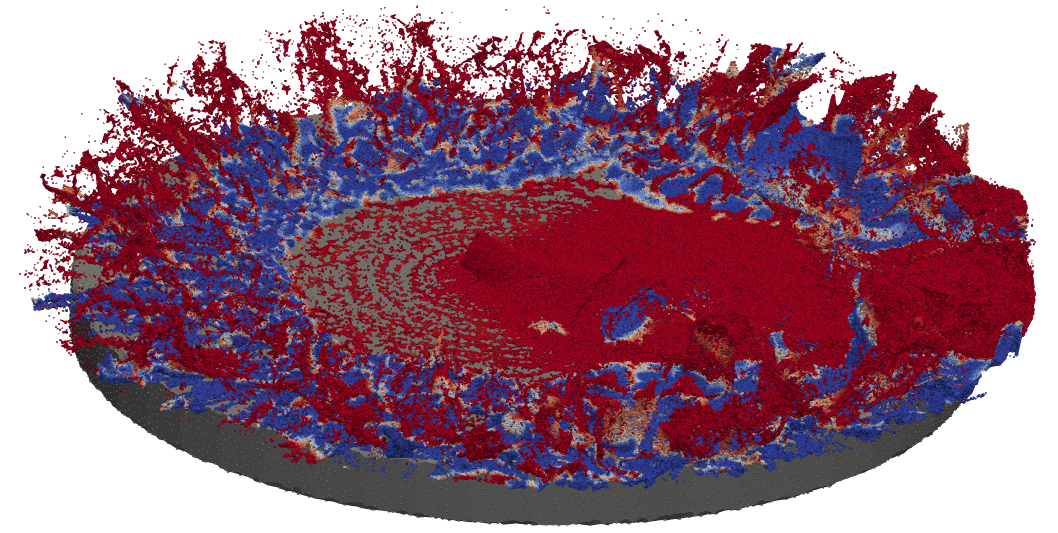}
  \includegraphics[width=0.49\textwidth]{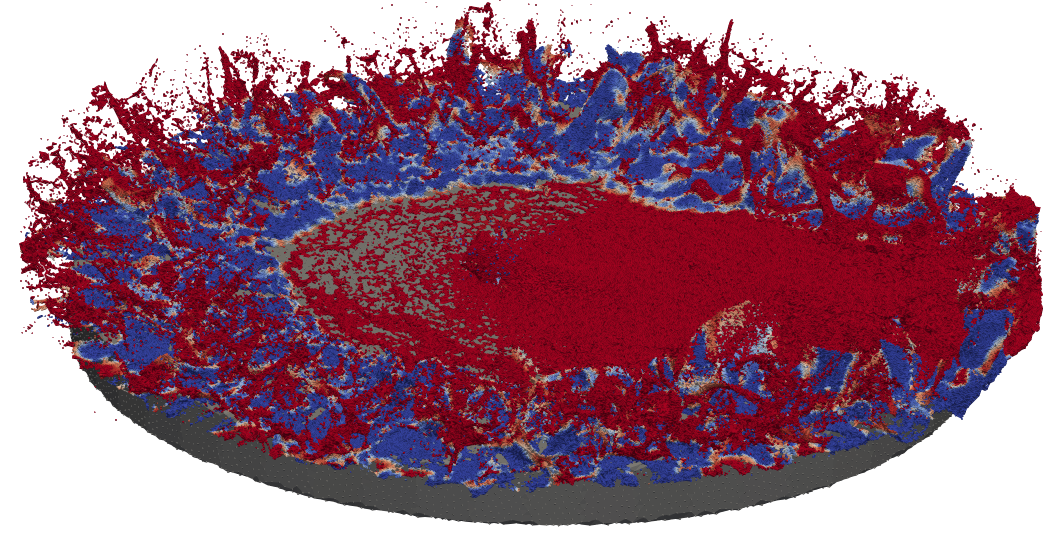}\\   
	\caption{Fluid jet impinging on a hot plate: Top view of the evolution of the boiling process at different time steps (see Figure~\ref{Fig:JetImpinge} for a side view). The color indicates the density. Top row: initial impingement and formation of the thin fluid layer. Middle row: Start of vaporization, and the jet pushing away the vaporization regions further. Bottom row: Formation of a temporally stable ring of vaporization around the main impingement location.}
  \label{Fig:JetImpinge_WithVaporisation}%
\end{figure}

The resultant vaporization process is shown in Figure~\ref{Fig:JetImpinge_WithVaporisation}. As the fluid jet impinges on the hot plate, it forms thin liquid layers on the solid surface. These are slowly heated by heat transferred from the solid. Since the inflow of the jet has a constant temperature, the thinnest regions of the fluid film which are further away from the jet impinging location reach boiling point and vaporize. As the vaporization starts and the hot plate is cooled down, the impinging jet pushes the initial vaporization regions further away. Figure~\ref{Fig:JetImpinge_WithVaporisation} also shows the vapor phase rising as a result of gravity. These observations qualitatively match experimental observations.

\begin{figure}
	\centering
	\begin{subfigure}[t]{0.45\textwidth}
		\centering
	    \begin{tikzpicture}
	    	\tiny
	    	\begin{axis}[
	    			xmin = 0, xmax = 0.8, ymin = 369, ymax = 401, xtick distance = 0.1, ytick distance = 4, grid = both, minor tick num = 1, major grid style = {lightgray}, minor grid style = {lightgray!25},
	    			width = \textwidth, height = \textwidth, xlabel = {time in \unit{\second}}, ylabel = {temperature in \unit{\degreeCelsius}}, legend pos=south west]
	    			\addplot[ only marks, mark=x,mark size = .8pt, blue 
	    			] file[skip first] {./Figures/T_over_time_Experiment.dat};
	    			\addlegendentry{experiment}
	    			\addplot[smooth, black, 
	    			] file[skip first] {./Figures/T_over_time_Simulation.dat};
	    			\addlegendentry{simulation},
	    	\end{axis}
	    \end{tikzpicture}
       	\subcaption*{(a)}
        \label{Fig:JetCompareExpA}
    \end{subfigure}\quad
    \begin{subfigure}[t]{0.45\textwidth}
    	\centering
    	\includegraphics[width=\textwidth]{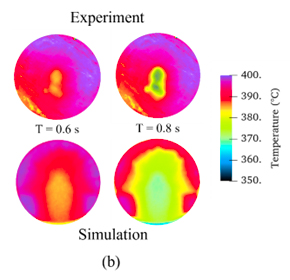}
    	\subcaption*{(b)}
    	\label{Fig:JetCompareExpB}
    \end{subfigure}\quad

    \caption{(a) Time history of the temperature in the center of the bottom solid surface. (b) Comparison of temperature at the bottom solid surface between experiment (top row) and simulation (bottom row) for two time instants.}
	\label{Fig:JetCompareExp}%
	
\end{figure}
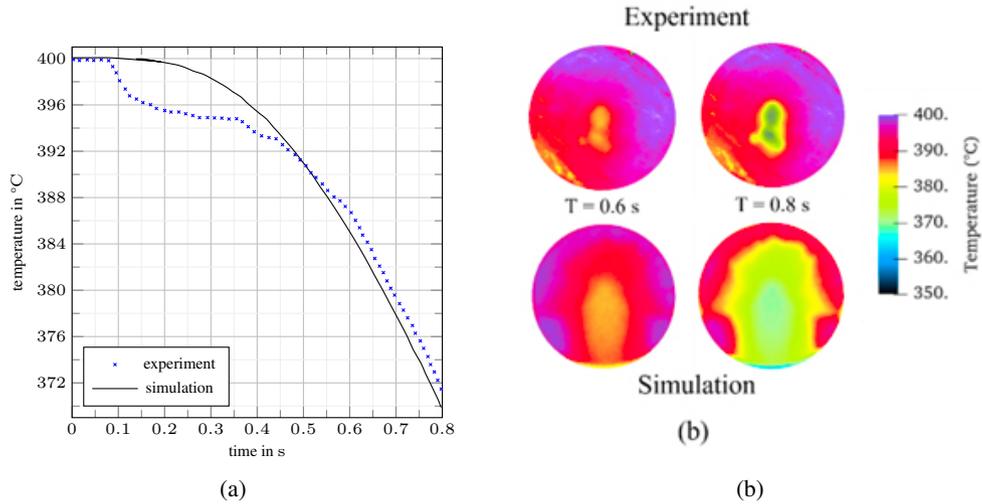

Figure~\ref{Fig:JetCompareExp}(a) shows a comparison between experimental and numerical results for the time history of the temperature at the center point of the rear solid surface. In the experiments \cite{Nabbout2022}, there is an oscillation initially when the temperature drop begins. This oscillation then stabilizes and both curves follow the same trend. In Figure~\ref{Fig:JetCompareExp}(b) the temperature distribution on the rear surface is displayed for two time instants in comparison between simulation and experiment. Compared to the experiment the wetting area is qualitatively the same. However, in the experiment, there is an asymmetry and the growth of the wetting area is more pronounced. This shows some differences in the local distribution, which indicate the potential for improvement for example by more detailed modeling of the heat transfer coefficient.

\section{Conclusion}
\label{sec:Conclusion}

We presented a novel one-fluid approach to simulating phase change processes, with a focus on fluid-to-vapor phase change at the boiling point. While conventional approaches to this problem explicitly model the interface between the fluid and vapor phases, either by a sharp interface or a smooth interface approach, our proposed method does not model the interface at all. The phase change process is modeled only through varying material properties, with the latent heat of phase change smeared out over a small temperature range around the boiling point. This enables us to simulate phase changes in complex flow domains, such as vaporization at free boundaries, which is extremely challenging with conventional interface-based methods. Our numerical results show the applicability of the method to such scenarios. 

Simulations were done using a Lagrangian meshfree approach using a strong form generalized finite difference method to approximate spatial derivatives. To capture the large gradients in the material properties, which is often challenging using a strong form approach, we enrich the test functions space to include discontinuous functions.  

While the numerical results suggest that one-fluid phase change modeling is very promising, they also raise several questions which need to be investigated. One such question is the optimal choice of the pressure interpolation in the phase change region. Another open question concerns the appropriate modeling of the very thin vapor layers and using an adaptive refinement based on the thickness of the vapor layers. Several possible improvements of the model proposed could be considered. This includes a more detailed modeling for the heat transfer coefficient by data-driven or multi-scale modeling \cite{Sato2015, Son1999, Wayner1976}, the incorporation of a turbulent thermal conductivity, as well as a modeling of local time scales for vaporization rates which could, for example, describe a pressure dependency.


Numerical results for the vaporization when a jet with free boundaries impinges on a hot surface, including comparisons with experiments, suggest suitability for the present method to model the vaporization of cutting fluid in metal cutting applications. Our future work is directed towards the application of the newly developed model to vaporization during metal cutting, with more detailed validations against experiments. First simulations have also shown the ability to simulate the Leidenfrost effect. A planned extension of this work is to thoroughly examine and validate the modeling of the Leidenfrost effect using the presented one-fluid phase transition model. 


\section*{Acknowledgements}
All the authors would like to acknowledge support from the Deutsche Forschungsgemeinschaft (DFG) under the priority program SPP 2231 ``FLUSIMPRO", project number 439626733. 
Pratik Suchde would also like to acknowledge partial support from  the  European  Union’s  Horizon  2020 research  and  innovation program under the Marie Skłodowska-Curie Actions grant agreement No. 892761 ``SURFING". 


\bibliographystyle{abbrv}




\end{document}